\documentclass[a4paper,fleqn,usenatbib]{mnras}

\usepackage{newtxtext,newtxmath}

\usepackage[T1]{fontenc}
\usepackage{ae,aecompl}

\usepackage{graphicx}	
\usepackage{amsmath}	
\usepackage{amssymb}	

\usepackage{mathrsfs}
\usepackage[flushleft]{threeparttable}

\title[The age of 47\,Tuc]{The age of 47\,Tuc from self-consistent isochrone fits to colour-magnitude diagrams and the eclipsing member V69}

\author[K. Brogaard et al.]{K. Brogaard,$^{1}$\thanks{E-mail: kfb@phys.au.dk}
D. A. VandenBerg,$^{2}$
L. R. Bedin,$^{3}$
A. P. Milone,$^{4}$
\newauthor 
A. Thygesen,$^{5}$
and F. Grundahl$^{1}$
\\
$^{1}$Stellar Astrophysics Centre, Department of Physics and Astronomy, Aarhus University, Ny Munkegade 120, 8000 Aarhus C, Denmark\\
$^{2}$Department of Physics and Astronomy, University of Victoria, P.O. Box 1700 STN CSC, Victoria, B.C., V8W 2Y2, Canada\\
$^{3}$INAF-Osservatorio Astronomico di Padova, Vicolo dell'Osservatorio 5, I-35122 Padova, Italy\\
$^{4}$Research School of Astronomy and Astrophysics, The Australian National University, Cotter Road, Weston, ACT, 2611, Australia\\
$^{5}$California Institute of Technology, 1200 E. California Blvd., MC 249-17, Pasadena, CA, 91125
}

\date{Accepted XXX. Received YYY; in original form ZZZ}

\pubyear{2016}

\begin{document}
\label{firstpage}
\pagerange{\pageref{firstpage}--\pageref{lastpage}}
\maketitle

\begin{abstract}

Our aim is to derive a self-consistent age, distance and composition for the globular cluster 47\,Tucanae (47\,Tuc; NGC104).
First, we reevaluate the reddening towards the cluster resulting in a nominal $E(B-V)=0.03\pm0.01$ as the best estimate. The $T_{\rm eff}$ of the components of the eclipsing binary member V69 is found to be $5900\pm72$ K from both photometric and spectroscopic evidence. This yields a true distance modulus $(m-M)_0=13.21\pm0.06$(random)$\pm0.03$(systematic) to 47\,Tuc when combined with existing measurements of V69 radii and luminosity ratio.
We then present a new completely self-consistent isochrone fitting method to ground based and {\it HST} cluster colour-magnitude diagrams and the eclipsing binary member V69.
The analysis suggests that the composition of V69, and by extension one of the populations of 47\,Tuc, is given by [Fe/H]$\sim-0.70$, [O/Fe]$\sim+0.60$, and $Y\sim0.250$ on the solar abundance scale of Asplund, Grevesse \& Sauval. However, this depends on the accuracy of the model $T_{\rm eff}$ scale which is 50-75 K cooler than our best estimate but within measurement uncertainties.
Our best estimate of the age of 47 Tuc is 11.8 Gyr, with firm ($3 \sigma$) 
lower and upper limits of 10.4 and 13.4 Gyr, respectively, in satisfactory
agreement with the age derived from the white dwarf cooling sequence {\it if
our determination of the distance modulus is adopted}. 

\end{abstract}

\begin{keywords}
globular clusters: individual: 47\,Tuc (NGC104) -- binaries: eclipsing -- Hertzsprung-Russel and colour-magnitude diagrams
\end{keywords}



\section{Introduction}



The ages of Galactic globular clusters (GCs) provide insights into the formation of
the Milky Way if they are determined to sufficient accuracy - whether in a
relative or absolute sense. The uncertainties associated with relative ages
are smaller than those for absolute ages because the former are derived from
cluster-to-cluster differences in the morphologies of their color-magnitude
diagrams (CMDs) and in their chemical properties. Absolute ages, on the other
hand, depend sensitively on the absolute distances and chemical abundances
(particularly of CNO). The study of a large sample of globular clusters by\defcitealias{VandenBerg2013}{VBLC13}\citet[hereafter VBLC13]{VandenBerg2013} and \cite{Leaman2013} provided very precise estimates of their ages. The relative ages implied by those investigations are expected to be accurate to within $\sim \pm 0.25$--0.5
Gyr, which is roughly one-half of the uncertainties associated with their
absolute ages.

Those studies relied on matching the observed horizontal branches with theoretical
zero age horizontal branch (ZAHB) loci to obtain the apparent cluster distances
--- an essential part of the age estimates. At low metallicities, it is
relatively straightforward to match a computed ZAHB to the lower bound of
the distribution of HB stars in a given GC. However, for the more metal-rich
clusters, this procedure is problematic because the HB is in fact not
horizontal, but bends towards brighter magnitudes at the red end.
To circumvent this problem, 47\,Tuc was used as an anchor-point for the
metal-rich globular clusters by adopting for this system the distance modulus
derived from the eclipsing cluster member V69 measured by \cite{Thompson2010}.
\citetalias{VandenBerg2013} found that the age of 47 Tuc obtained from the mass-radius diagram of the V69 components, 11.0 Gyr, was 0.75 Gyr less than the age that they derived from fits of isochrones to {\it HST} photometry of the cluster if the models assume $Y = 0.257$, [Fe/H]$ = -0.76$ \citep{Carretta2009}, and $[\alpha/\rm{Fe}]=+0.46$ relative to the solar abundances by \cite{Grevesse1998}.  Although they showed that this discrepancy could depend on the adopted helium abundance and/or [Fe/H] and $[\alpha/\rm{Fe}]$, they decided to postpone a more detailed analysis to a subsequent investigation when this matter could be examined more thoroughly. The present paper is the planned follow-up study.

In the meantime the true distance modulus of 47\,Tuc as inferred from the
white dwarf (WD) population of the cluster was reported by \cite{Woodley2012} to be $(m-M)_0=13.36\pm0.08$ mag while \cite{Watkins2015} reported a dynamical estimate of the true distance corresponding to $(m-M)_0=13.09\pm0.04$. 

\cite{Hansen2013} utilized the white dwarf cooling sequence to obtain an age
estimate as young as $9.9\pm0.7$ Gyr (95\% confidence) for 47\,Tuc for a
corresponding true distance modulus of $13.32\pm0.09$. The same authors reanalyzed the eclipsing binary V69 yielding an age of $10.39\pm0.54$
Gyr (systematic error only).
\cite{GarciaBerro2014} obtained an age close to 12 Gyr from the white dwarfs
for a corresponding true distance modulus of 13.2 mag (Garcia-Berro, private
comm.). Thus it would seem that the distance modulus and age of 47\,Tuc are
not well established and that further investigation is warranted --- not only
for the sake of understanding the cluster itself, but also the other metal-rich
GCs for which 47\,Tuc serves as an anchor point in the study of GC ages
by \citetalias{VandenBerg2013}. 

In this paper we derive the distance and age of 47\,Tuc from self-consistent isochrone fits to colour-magnitude diagrams (CMDs) and the parameters of the components of the eclipsing member V69. In section 2 we describe the data used for the analysis and derive the effective temperatures and the distance moduli of the V69 components. Section 3 explains the isochrone fitting procedure. Results are presented and discussed in Section 4, starting with assumptions of perfect models and a simple stellar population (Sects. 4-4.2), extending to potential consequences of uncertain model assumptions (Sect. 4.3), and multiple populations (Sect. 4.4). A comparison to the age derived from the white dwarf cooling sequence is made in Sect. 5 before addressing potential consequences for relative ages of GCs in Sect. 6. Section 7 contains summary, conclusions and outlook. 

\section{Parameters of the eclipsing binary V69 and 47\,Tuc}

\subsection{V69}
\label{sec:V69}
V69 is a detached eclipsing binary (P=29.54 days) member with components very near the turn-off of 47\,Tuc, discovered by \cite{Weldrake2004} and measured and analysed by \cite{Thompson2010}. We adopted the masses and radii of V69 obtained by \cite{Thompson2010}, as well as their luminosity ratio and their finding that the two components have almost identical effective temperatures. 
Due to the near-identical temperatures, the luminosity ratio does not vary strongly with wavelength and is thus not expected to be significantly different between filters with nearby effective wavelengths. Based on this fact we adopted the luminosity ratio of $\frac{L_s}{L_p}=0.792\pm0.011$ from the $V$-band \citep{Thompson2010} to calculate the individual component magnitudes not only in $V$ but also in $F606W$. Table~\ref{table:data} contains the adopted and derived magnitudes and parameters for V69.

\subsection{Photometry}

   \begin{figure*}
   \centering
   \includegraphics[width=17cm]{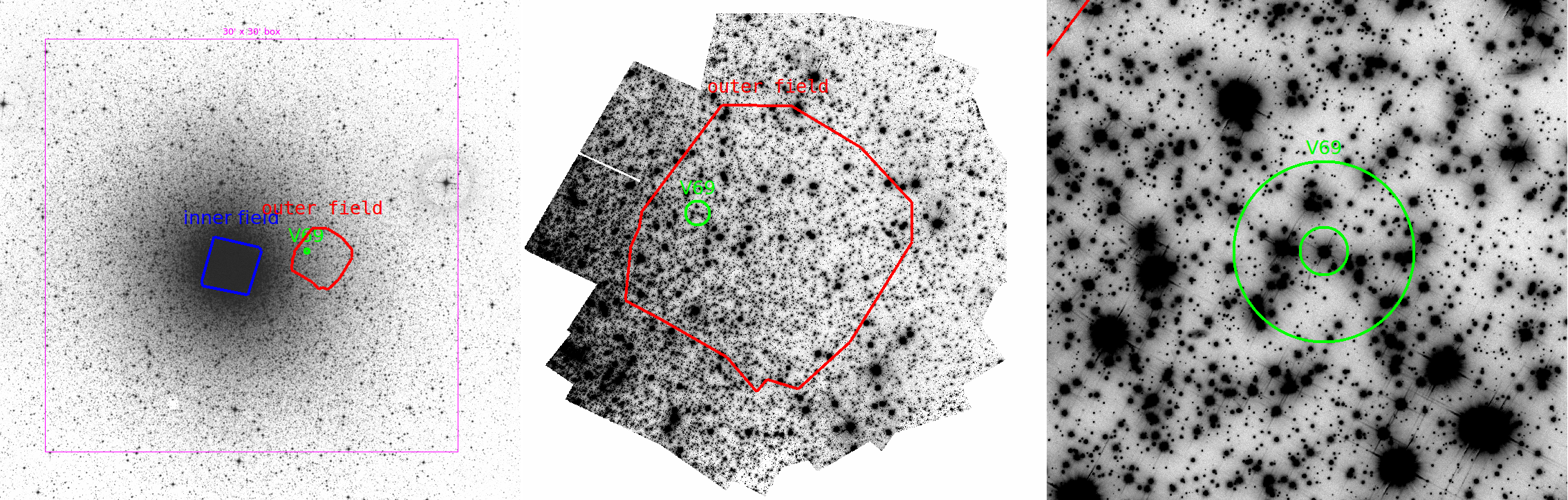}
   \caption{{\it (Left):} overimposed to the Digital Sky Survey (DSS) we show the footprints of the {\it HST} fields analyzed in this work. The inner field is indicated in blue, outer field in red (see text), while a 30$^\prime$$\times$30$^\prime$ box is given for reference. North is up, East is Left. The position of V69 is indicated in green. {\it (Middle):} same footprint overimposed to a stack image in filter F814W for the {\it HST} outer region. The red footprint has a size of about 3$^\prime$$\times$3$^\prime$. {\it (Right):} a zoom-in around V69, meant to be a finding chart. The largest of the green circles has a radius of 11 arcsec, while the field of view is about 1$^\prime$$\times$1$^\prime$.}
              \label{fig:fields}%
    \end{figure*}

For the analysis we made use of the parameters of V69 in conjunction with CMDs of 47\,Tuc constructed from the ground based photometry by \cite{Bergbusch2009} and {\it HST} photometry produced for this work from an {\it HST} archival pointing where V69 is located. 

\subsubsection{$BVI_C$}

We employed the $BVI_C$ photometry presented by \cite{Bergbusch2009}. We were
unable to locate V69 in the photometry using coordinates from either
\cite{Weldrake2004} or \cite{Kaluzny1998} but were kindly informed (J. Kaluzny, private comm.) that the online catalog of the latter reference gives J1950
coordinates while the README file incorrectly says J2000, and that the J2000
coordinates of V69 are $(\alpha,\delta)=$ (00:22:53.65,-72:03:46.64). We did not find the reason for the discrepant
coordinates of Weldrake et al. 

With the correct coordinates, we identified V69 as star number 44164 in the
photometry of \cite{Bergbusch2009}. However, the magnitudes of V69 were fainter by 0.041 mag in $V$ and by
0.027 mag in $B$ compared to those reported by \cite{Thompson2010}, although these authors report
a mean offset of their photometry of only 0.004 mag compared to 20 secondary
standard stars in 47\,Tuc from the Stetson catalog (Stetson 2000) that we
verified had an even smaller offset compared to Bergbusch \& Stetson (2009).
An inquiry about the discrepancy lead to the discovery that some observations
were taken during eclipse (P. Stetson, private comm.). After removing these
measurements from the light curve of V69 kindly provided by P. Stetson, we
obtained new out-of-eclipse magnitudes of V69 on the scale of Bergbusch \&
Stetson (2009) by taking weighted averages of the remaining measurements in
each filter. These values, which we report in Table~\ref{table:data}, are
in close agreement (to within 0.01 mag) with Thompson et al. (2010) for $B$
and $V$ and provide also a precise and self-consistent value for $I_C$,
which was not measured directly by \cite{Thompson2010}.

\begin{table*}
\centering
\caption{\label{table:data}Measurements of V69}
\begin{threeparttable}
\begin{tabular}{lccc}
\hline\hline
Quantity&System&Primary&Secondary\\
\hline
Period (days)   &	$29.53975(1)$\tnote{1} & (...) & (...)\\
RA  (J2000)\tnote{9} & $00:22:53.67$ & (...) & (...)\\
DEC (J2000)\tnote{9} & $-72:03:46.7$ & (...) & (...)\\
Mass$(M_{\odot})$&	(...)	&$0.8762(48)$\tnote{1}	&$0.8588(60)$\tnote{1}\\
Radius$(R_{\odot})$&	(...)	&$1.3148(51)$\tnote{1}	&$1.1616(62)$\tnote{1}\\
log$g$          &       (...)   &$4.143(3)$\tnote{1}    &$4.242(3)$\tnote{1}\\
$L_s/L_p$	&	$0.792(11)$\tnote{1}&(...)&(...)\\
$B$		&	$17.390(1)$\tnote{2}	&	(...)		&	(...)\\		
$V$		&	$16.826(1)$\tnote{2}	&$17.459$\tnote{3}&$17.713$\tnote{3}\\
$I_c$		&	$16.130(2)$\tnote{2}	&	(...)		&	(...)\\
$m_{F606W}$	&	$16.6133(8)$	&$17.2466$\tnote{3}&$17.4998$\tnote{3}\\	
$m_{F814W}$	&	$16.0804(11)$	&	(...)		&	(...)\\
$B-V$		&	$0.564(1)$	&	(...)		&	(...)\\
$V-I_c$		&	$0.696(2)$	&	(...)		&	(...)\\
$m_{F606W}-m_{F814W}$	&	$0.5329(13)$	&	(...)		&	(...)\\
\hline
\multicolumn{4}{c}{Assuming a nominal reddening of $E(B-V)=0.030$\tnote{4}:}\\
$(B-V)_0$		&	$0.536$		&	(...)		&	(...)\\
$(V-I)_0$		&	$0.660$		&	(...)		&	(...)\\
$(m_{F606W}-m_{F814W})_0$		&	$0.5033$	&	(...)		&	(...)\\
$T_{\rm eff} (K)$\tnote{5}	&	$5900(72)$	&$5900(72)$		&$5900(72)$\\
$BC_V$			&$-0.099(8)$	&$-0.099(8)$	&$-0.099(8)$	\\
$BC_{F606W}$		&$+0.080(8)$	&$+0.080(8)$	&$+0.080(8)$	\\
$(m-M)_V$\tnote{6}		&$13.30(6)$	&$13.30(6)$	&$13.29(6)$	\\
$(m-M)_{F606W}$\tnote{6}		&$13.26(6)$	&$13.27(6)$	&$13.25(6)$	\\
$(V-M_V)_0$\tnote{7}		&$13.20(6)$	&$13.21(6)$	&$13.19(6)$	\\
$(m_{F606W}-M_{F606W})_0$\tnote{7}		&$13.17(6)$	&$13.18(6)$	&$13.17(6)$	\\
distance (pc)\tnote{8}		& $4365\pm120$ & & \\	
\hline	
\multicolumn{4}{c}{Assuming a nominal reddening of $E(B-V)=0.040$\tnote{4}:}\\
$(B-V)_0$		&	$0.527$		&	(...)		&	(...)\\
$(V-I)_0$		&	$0.648$		&	(...)		&	(...)\\
$(m_{F606W}-m{F814W})_0$		&	$0.4934$	&	(...)		&	(...)\\
$T_{\rm eff} (K)$\tnote{5}	&	$5950(72)$	&$5950(72)$		&$5950(72)$\\
$BC_V$			&$-0.094(8)$	&$-0.094(8)$	&$-0.094(8)$	\\
$BC_{F606W}$		&$+0.081(8)$	&$+0.081(8)$	&$+0.081(8)$	\\
$(m-M)_V$\tnote{6}		&$13.34(6)$	&$13.34(6)$	&$13.33(6)$	\\
$(m-M)_{F606W}$\tnote{6}		&$13.30(6)$	&$13.31(6)$	&$13.29(6)$	\\
$(V-M_V)_0$\tnote{7}		&$13.21(6)$	&$13.22(6)$	&$13.20(6)$	\\
$(m_{F606W}-M_{F606W})_0$\tnote{7}		&$13.18(6)$	&$13.19(6)$	&$13.18(6)$	\\
distance (pc)\tnote{8}		& $4405\pm120$ & & \\	
\hline
\end{tabular}

\begin{tablenotes}
		\scriptsize
\item[1] Adopted from \cite{Thompson2010}.
\item[2] Recalculated magnitudes on the scale of \cite{Bergbusch2009}, excluding observations taken in eclipse. See text for details.
\item[3] Calculated from our magnitudes for the total light of V69 using the V-band light ratio from \cite{Thompson2010}.
\item[4] We make use of Table A1 in \cite{Casagrande2014} and adopt $T_{\rm eff}=5900K$ and $[\rm Fe/H]=-0.76$ to calculate $R_x$ values according to the formula in their table. We get $R_B=4.047, R_V=3.127, R_I=1.884, R_{606}=2.870, R_{814}=1.882$, from which we calculate factors of 0.920,1.243, and 0.987 to multiply the nominal $E(B-V)$ in order to get $E(B-V)$, $E(V-I)$, and $E(606-814)$ for a spectral type corresponding to V69.
\item[5] The mean value of effective temperature calculated from the three colours using the calibration of \cite{Casagrande2014} and three values of $[\rm Fe/H], -0.64,-0.70$, and $-0.76$. An uncertainty of $72 K$ is adopted to allow for uncertainties in the observed colours, the adopted reddening, colour-to-colour differences, and potential systematics the colour-temperature calibration.
\item[6] Calculated using the formalism in \cite{Torres2010} and $BC_{V,\odot}=-0.068$. The number for 'system' is the mean of the results for each component.
\item[7] The true distance modulus as calculated from $V$ and $F606W$, respectively. Note that in principle the true distance modulus should be the same regardless of the filter used to measure it, and the difference between the two must be due to zero-point errors, either in one or both photometries, or in the bolometric corrections.
\item[8] Using the $V$ results.
\item[9] Derived in this paper on the Gaia system.
\end{tablenotes}
\end{threeparttable}
\end{table*}

\subsection{\textit{HST} ACS/WFC photometry}
\label{sec:hst}
We made use of photometry obtained with the {\it Hubble Space Telescope} ({\it HST}), and in particular 
 images collected  with the \textit{Wide Field  Channel} (WFC)  of the
\textit{Advanced Camera  for Surveys}  (ACS). Photometry is extracted
from two fields mapping two different regions of 47\,Tuc.

The location of both fields with respect to 47\,Tuc and V69 are shown in the left panel of
Fig.~\ref{fig:fields} superimposed to a DSS image. A zoom-in of the outer field is shown in the middle panel 
superimposed on the stacked image that was made and released by Anderson
et al.\ (2007; \texttt{outpix.9Kx9K.fits})\footnote{http://www.stsci.edu/$\sim$jayander/47TUC\_CAL/}. On the right panel of Fig.~\ref{fig:fields} a finding chart of V69 is given.

All the \textit{HST} photometry considered for this work is obtained with the
ACS/WFC filters $F606W$ and $F814W$.

\subsubsection{Outer Field}

The first field, hereafter referred to as the \textit{outer} field, has been reduced
specifically for this work. This field includes the eclipsing binary V69 and is therefore crucial for our analysis. It is located about 7 arcmin from the
center of 47\,Tuc, and was adopted by STScI as a main calibration field for
several purposes (CTE  monitoring, geometric  distortion, PSFs, etc.). A
massive number of images is available for almost any
possible exposure length and many filters; they were used in many
Instrument Science Reports for the ACS/WFC (eg., 2013-03, and references therein).

As we are interested in images where V69 is not severely saturated, we
restricted ourselves only to those with exposure times shorter than about
360\,s.  This resulted in 228 images in $F606W$ and 59 in $F814W$.
We downloaded from the MAST\footnote{http://archive.stsci.edu/hst/search.php} archive the
images corrected for CTE with the algorithm described in Anderson \& Bedin
(2010). 
We extracted from each image the
photometry and astrometry of individual point sources as described in
great detail in Anderson \& King (2006).  The method is essentially a
PSF-fitting, where PSFs are optimized for the ACS/WFC undersampled
images. The software is named img2xym\_WFC.09x10 and is also described and made
publicly available by \cite{Anderson2006}.\footnote{See http://www.stsci.edu/$\sim$jayander/ACSWFC\_PSFs/ for the code as well as library PSFs.}

Next, we linked all the extracted individual-image catalogs into a
common reference frame.  The adopted reference frame is the the
\texttt{MEMBER.RIGID.XYM} which is described and released as part of
\cite{Anderson2007}.\footnote{ 
The reference frame MEMBER.RIGID.XYM was obtained by Anderson (2007) and made publicly available as
part of the Instrumental Science Report ACS-WFC 007-008.
The MEMBER.RIGID.XYM catalog is based on 193 F606W observations of the outer 
calibration field in 47\,Tuc at various positions and roll
angles. For each exposure fluxes and positions were measured for all the stars in the image using
the program img2xym\_WFC.09x10 (web-link in footnote 3), with the positions corrected for distortion
using the solution available at http://www.stsci.edu/$\sim$jayander/GCLIB/ \citep{Anderson2006}. 
A collating routine was then used to
generate a rough master frame for the entire field, and cross-identify
and average the star positions and fluxes measured in the different exposures. 
Anderson (2007) found that the linear skew terms of the distortion
solution of ACS/WFC have been changing at a rate of about 0.04 pixel
per year. He provided a simple correction, and
showed that the remaining solution is globally accurate to 0.02 pixel,
i.e., 1 mas. The corrected positions for cluster member stars in the improved
master frame are those found in the file MEMBER.RIGID.XYM that
can be used as an accurate frame of reference. More details are available at \hfill\break http://www.stsci.edu/hst/acs/documents/isrs/isr0708.pdf
}

Then, for each filter, we combined all the individual catalogs to
obtain clipped-mean magnitudes around the median, and keeping only stars detected
at least 70 times out of the 228 available $F606W$ images, and 22 times
out of the $59$ for F814W images. The catalog for each filter is an
average of several epochs spanning several years \citep{Ubeda2013},
nevertheless the internal motion of 47\,Tuc corresponds to just a few
hundred pixels over a ten-year baseline.

To reject most of the spurious detections, mismatches, and poorly
measured stars, we imposed a consistency of 0.1 pixels (about 5\,mas)
between the average positions measured in the $F606W$ and the $F814W$
images.  The resulting list contains very solid photometric
detections.

\subsubsection{Inner Field}

The second field, hereafter the \textit{inner}  field, is centered  on the
core  of 47\,Tuc,  and  was observed  as  part of  the  ACS Survey  of
Galactic Globular  clusters (GO-10775,  PI: Sarajedini,  Sarajedini et
al.\ 2007) and described in detail by \cite{Anderson2008}. 
We also re-reduced this field independently, using
the software by Anderson \& King (2006), see next section.

\subsubsection{Calibration}
\label{sec:calibration}
The calibration of both the inner and outer field to the VEGAMAG standard
system was done using the procedure detailed in \cite{Bedin2005}.

\begin{figure}
   \centering
   \includegraphics[width=8.6cm]{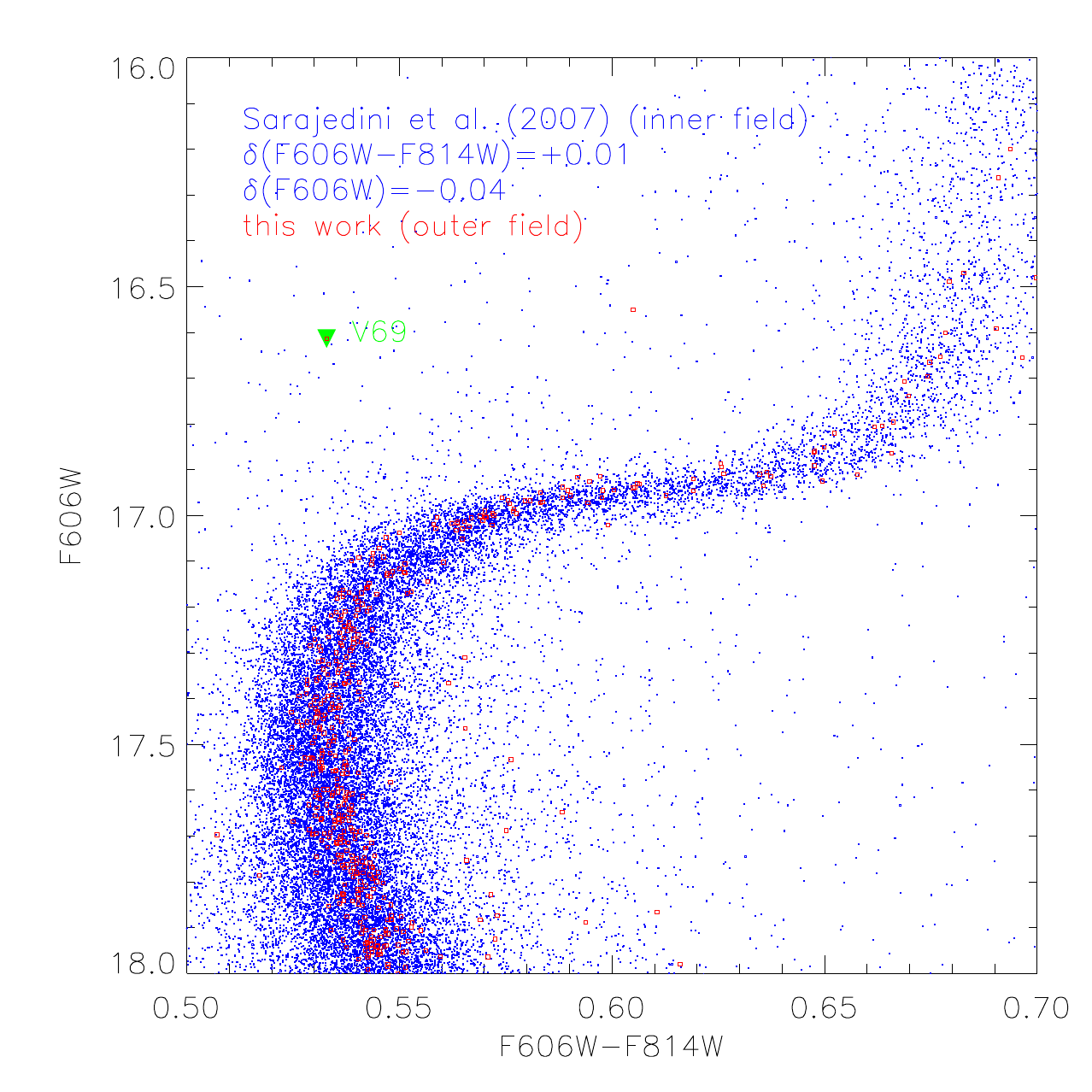}
   \caption{Comparison of the cluster sequences of the inner and outer fields
of 47\,Tuc as defined in Fig.~\ref{fig:fields}. The inner field photometry,
which is from Sarajedini et al. (2007), has been shifted in colour and
magnitude by the indicated $\delta$ amounts relative to the outer field.
These shifts have been calculated using the same procedure that was described in
Sect.~\ref{sec:procedure} using the point on the subgiant branch 0.035 mag
redder than the turn-off colour in order to avoid subjective judgments. The
relative shift is, however, sensitive the quality criteria adopted for stars
included in the inner field. Here we have used all stars; but selecting
instead only those stars with $\sigma_{(F606W-F814W)} \leq 0.005$ results in
$\delta(F606W)=-0.06$.}
              \label{fig:io}%
    \end{figure}

\begin{figure}
   \centering
   \includegraphics[width=8.6cm]{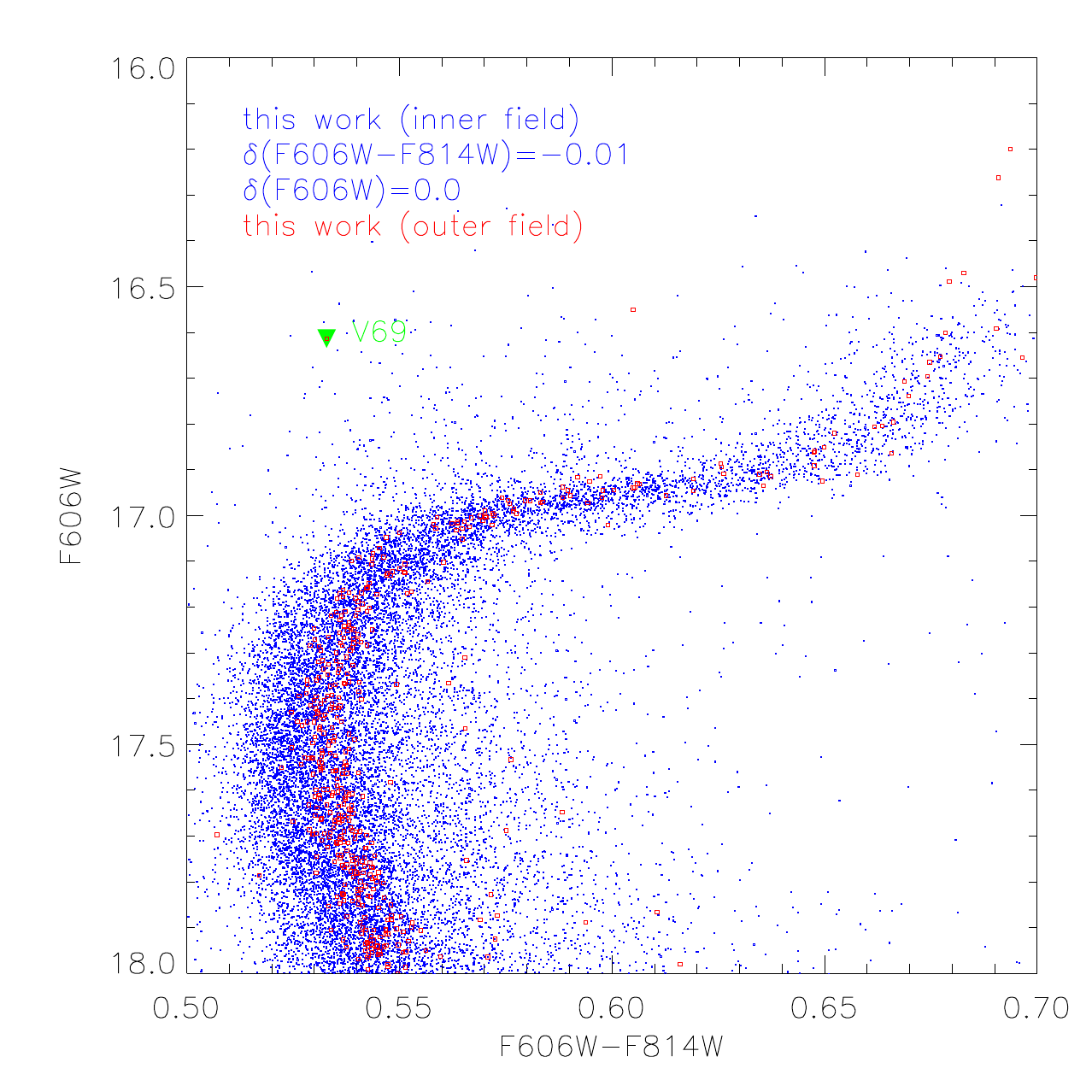}
   \caption{Similar to Fig.~\ref{fig:io} except that the inner field photometry has been derived as part of this work using the exact same procedures as for the outer field.}
              \label{fig:iotw1}%
    \end{figure}

For the inner field we first adopted the calibrated and publicly
released photometric catalog \citep{Sarajedini2007, Anderson2008},
which was also analysed by \cite{Milone2012}. However, when aligning
the CMDs of the outer and inner fields using the procedure described
in Sect.~\ref{sec:procedure} that determines the turn-off colour and
the magnitude of the point on the subgiant branch which is 0.035 mag
redder, we found rather large zero-point differences between the inner
and outer field, see Fig~\ref{fig:io}.  We therefore decided to redo
the photometry of the inner field, so that both fields are reduced in
exactly the same way, i.e.
using exactly the same software by Anderson \& King (2006) as used for the outer fields
in Section\,2.3.1, which is more accurate and precise for relatively isolated bright stars,
and most importantly now self-consistent between the inner and outer fields.
Doing that, indeed, 
we find very good consistency between the two fields, as seen in
Fig.~\ref{fig:iotw1}. The remaining offsets are within the expected
uncertainties of the reduction and calibration procedures and other
causes, such as: potential biases due to different amounts of crowding
in the two fields, possible differential reddening effects, and PSF
modelling. The photometry catalogs of both fields are published along
with the
paper\footnote{http://web.oapd.inaf.it/bedin/files/PAPERs\_eMATERIALs/47Tuc}.

%

%

\subsection{Reddening}

To calculate the photometric $T_{\rm eff}$ values, we needed a reddening estimate. The reddening of 47\,Tuc is $E(B-V)=0.04$ according to the 2010 revision of the Harris catalog \citep{Harris1996}. 
We explored the literature in search of the derivation and accuracy of this number. \cite{Hesser1987} state that the best evidence favours $E(B-V)=0.04\pm0.01$ for 47\,Tuc citing 
\cite{Crawford1975}, \cite{Hesser1976}, and \cite{Lee1977}. However,
\cite{Hesser1976} "recommend using $E(B-V) = 0.03$ for 47\,Tuc in the future".
Considering the references used by Hesser et al. (1987) to favour $E(B-V)=0.04$,
both Crawford \& Snowden (1975) and Hesser \& Philip (1976) obtained
$E(B-V)=0.03$, while Lee (1977) does not claim to derive a value but
only that results are consistent with $E(B-V)=0.04\pm0.01$ as derived by other researchers.
This is rather vague, and actually the two papers quoted by Lee (1976) do not
claim such a low uncertainty. On the contrary, they base their estimates on
means of measurements with a total range that is twice the mean
value \citep{Cannon1974, Hartwick1974}, suggesting that the uncertainty is
as large as the reddening itself. Thus it would appear that the number adopted
by Hesser in 1987, $E(B-V)=0.04\pm0.01$, which is likely the source of the
value in the Harris catalog, could as well have been $E(B-V)=0.03\pm0.01$. This
is far from a complete picture, but it seems that although some of the more
precise work from the seventies suggested $E(B-V)=0.03$, many researchers
continued to use the value of $E(B-V)=0.04$ long after the \cite{Schlegel1998} dust map value of $E(B-V)=0.032$ became available, likely because
the Harris catalog gives 0.04.

Much more recently, Schlafly et al. (2011) derived a reddening map based on
SDSS measurements of stars, giving $E(B-V)=0.028$ for 47\,Tuc. Although this is
14\% lower than the value derived by Schlegel et al.,
about half of this difference appears to be due to the
different spectral energy distributions (SEDs) of the objects that are used to derive
the extinction law. The reddening maps of Schlegel et al. are valid for galaxies
while the maps of Schlafly et al. are valid for stars with effective
temperatures around 7000 K (see the footnote of their Table 6).
As described by Casagrande \& VandenBerg (2014), the {\it nominal} reddening, defined as the reddening of a hot star with a flat SED, is larger than the reddening of the cooler turn-off stars, and for 47\,Tuc this difference is $\Delta E(B-V)=0.002$. We therefore adopt a best estimate nominal reddening of $E(B-V) = 0.03$ for 47\,Tuc so that the reddening at the turn-off colour is 0.028, in agreement with Schlafly et al. (2011) for stars of similar effective temperature. 
The very good agreement between the $E(B-V)$ values for 47\,Tuc from Schlegel
et al. and Schlafly et al. suggests that the higher value in the Harris catalog
is too high, which would not be surprising given that all of the
$E(B-V)=0.04$ estimates that we have been able to find in the literature
have rather much larger uncertainties. 

To also have our own independent reddening estimate we measured the
equivalent widths of the interstellar Na I D lines in the spectra of 47\,Tuc
giants from \cite{Thygesen2014} and used the calibration of \cite{Munari1997}
to transform these to $E(B-V)$. Synthetic spectra were used to remove the stellar lines. This resulted in $E(B-V)=0.020\pm0.002$ where the uncertainty is the error of the mean from 5 spectra. However,
this value is an extrapolation to lower values than those measured for any
star by \cite{Munari1997} and is therefore much more uncertain than the
random error and not useful for lending much more credibility to $E(B-V)=0.03$
than $E(B-V)=0.04$ although it is closer to the former value.

In order to examine the effects of the reddening uncertainty in the following
analysis, we adopt $\pm0.01$ as the uncertainty on $E(B-V)$ and
use two different estimates of $E(B-V)$ ($0.03$, which
corresponds to our best estimate, as justified above, and $0.04$, as listed
in the \cite{Harris1996} catalogue) when calculating the photometric
$T_{\rm eff}$ values. Results for $E(B-V)=0.02$ can be inferred by extrapolating from these two cases in Table~\ref{table:data}.

\subsection{$T_{\mathrm{eff}}$ of V69}
\label{sec:teff}
We used the calibration by Casagrande \& VandenBerg (2014) to deredden the
colours of V69 and transform them into effective temperatures. Since the
colours of the two components are very similar we assumed that one
$T_{\rm{eff}}$, as derived from the colours of the combined light, represents
both components to a good approximation. We adopted a mean from the three
colours $(B-V)$, $(V-I_C)$, and $(F660W-F814W)$, each calculated assuming three
different values of [Fe/H], $-0.76,-0.70$, and $-0.64$. These nine numbers
are very similar for a fixed reddening; all but two differ from the mean
$T_{\rm eff}$\ value by less than 20 K and none of them differ from the mean
value by more than 45 K. Assuming a nominal $E(B-V)=0.03$ the mean effective temperature is 5900 K and
assuming $E(B-V)=0.04$ the mean is 5950 K. If we use instead the empirical
calibration by \cite{Casagrande2010} we obtain a hotter temperature by just 7 K
from $V-I_C$ and by $42-58$ K (depending on the assumed reddening) from $B-V$.
Averaging results from the two filters yields a mean $T_{\mathrm{eff}}$ of $5933$ K for $E(B-V)=0.03$ and 5974 K for $E(B-V)=0.04$, just 33 K and 24 K hotter than using the \cite{Casagrande2014} calibration. This demonstrates very good agreement between the theoretical and empirical colour-$T_{\rm{eff}}$ relations. 

As argued above, we adopt the value of $T_{\mathrm{eff}}$ assuming $E(B-V) = 0.03$ as our best estimate.  That is, $5900$ K, as found from the color transformations given by \cite{Casagrande2014}, will be used throughout the following analysis for the sake of maximal self-consistency. 

\begin{table*}
\centering
\caption{Spectroscopic parameters of turn-off stars}

\begin{threeparttable}
\label{tab:TOTeff}

\begin{tabular}{lccccccc}
\hline\hline
ID\tnote{1}	& RA(2000)\tnote{1}	& DEC(2000)\tnote{1}	& $T_{\rm eff}$	& log$g$	& $\xi_t$		& $[\rm{FeI/H}]$	 & $[\rm{FeII/H}]$ \\

\hline
1081	& $00:21:03.82$ & $-72:06:57.74$ & $5980\pm162$	& $3.95\pm0.40$	& $1.60\pm0.24$	& $-0.94\pm0.23$ & $-0.93\pm0.19$ \\
1012	& $00:21:26.27$ & $-72:00:38.73$ & $5720\pm129$	& $3.65\pm0.40$	& $0.85\pm0.23$	& $-0.97\pm0.20$ & $-0.94\pm0.13$ \\
975	& $00:20:52.30$ & $-71:58:01.80$ & $5850\pm157$	& $3.65\pm0.40$	& $1.10\pm0.25$	& $-0.66\pm0.22$ & $-0.66\pm0.25$ \\
\hline

\end{tabular}
\begin{tablenotes}
		\item[1] IDs and coordinates from Carretta et al. (2004).
\end{tablenotes}
\end{threeparttable}

\end{table*}

We searched the literature for $T_{\rm eff}$ estimates that do not depend on
reddening. \cite{Carretta2004} found a spectroscopic estimate of $5832$\ K for
the mean spectrum of three turn-off stars. Their Fig. 1 suggests that these stars are significantly different from each other. We therefore reanalysed the spectra of these stars, which we obtained from the ESO archive, and found that only one of them, star 1081, seems to be a good representative of V69 when comparing the surface gravities. This is also implied by the CMD position of the three stars in Fig. 1 of \cite{Carretta2004}. We give our measured quantities for the three stars in Table~\ref{tab:TOTeff}. 
In view of the close similarity of their $V$-magnitudes and log$g$ values, V69 should have close to the same effective temperature as star 1081, $5980\pm162$, which is unfortunately rather uncertain due to the low S/N of the spectrum and thus consistent with both our estimates above.

\cite{Dobrovolskas2014} measured $T_{\rm eff}$ values from spectra of turn-off
stars by $H\alpha$ profile fitting. We found in their table A1 ten stars that
have log$g$, $V$-mag and $(B-V)$ very similar to the primary star of V69. We chose those that
have $V$-mag within 0.05 mag and log$g$ within 0.04 dex of the V69 primary component. The mean $T_{\rm eff}$ of these ten stars is
$5882\pm47 (\rm{RMS})$\ K, with the range spanning from $5780$\ K to $5968$\ K
without a correlation with $(B-V)$, suggesting either a random error larger
than measured or low quality photometry. If the errors were random, the
uncertainty on the mean would be $\sqrt{10}$ lower, yielding only $16$\ K, but
in this case none of the individual measurements would be within
$1\sigma$ of the true value. Therefore, systematic errors seem to dominate
these measurements, and the RMS of $47$ K is therefore adopted as the $1\sigma$ uncertainty of this spectroscopic estimate.

Our photometric estimate from above has uncertainty components arising from reddening, photometric measurements, and the colour-$T_{\rm eff}$ relations. As derived above, the uncertainty due to reddening is $50$ K and from the colour-colour differences we estimate conservatively a scatter of $50$ K, which averaging the three colours gives $28$ K. Ignoring for now the accuracy of the calibration and adding the two contributions in quadrature gives a $1\sigma$ estimate of $57$ K.

Both the photometric and spectroscopic $T_{\rm eff}$ estimates and their uncertainties are thus very similar and a mean of the two estimates has an uncertainty of $39\, K$. To this we add, {\it directly, not in quadrature}, another $33$ K such that a full $3\sigma$ uncertainty also accommodates a potential zero-point shift of the {\it observed} temperature scale by $100$ K, which seems difficult to rule out (see e.g. \citealt{Casagrande2010} and \citealt{Molenda2014}). 

Thus, based on the evidence presented we adopt a best estimate effective temperature of of $5900\pm72$ K for the V69 components.

\subsection{Distance}
\label{sec:distance}
We calculated the apparent and true distance moduli for
the $V$ and $F606W$ filters (see Table~\ref{table:data}) as follows. 
First, we used equation (10) of \cite{Torres2010} generalised to an arbitrary filter $X$ to calculate the absolute component magnitudes of V69:
\begin{equation}
M_X=-2.5\rm{log}\left(\frac{L}{L_\odot}\right)+V_\odot+31.572-\left(BC_X-BC_{V,\odot}\right)
\end{equation}

Here, $V_\odot=-26.76$ as recommended by \cite{Torres2010} and $BC_{V,\odot}=-0.068$ as obtained from the calibration of \cite{Casagrande2014}. The luminosities were obtained from the component radii and the $T_{\rm eff}$ as derived in the previous section, using $T_{\rm eff,\odot}=5777$K.
The apparent distance moduli were then obtained by combining the absolute component magnitudes with the corresponding apparent magnitudes calculated from our measured system magnitudes and luminosity ratio for the $V$-band (see Sects.~\ref{sec:V69}-\ref{sec:hst}). These were also translated to true distance moduli for two values of $E(B-V)$ using $R_X$ values from \cite{Casagrande2014} as detailed in the footnote of Table~\ref{table:data}.

Our derived distance modulus for
the $V$-band is in excellent agreement with that derived for V69 by
\cite{Thompson2010} if we adopt $E(B-V)=0.04$ as they did. For our preferred
reddening of $E(B-V)=0.03$, we get a lower number, $(m-M)_V=13.30\pm0.06$.
Converted to a distance this is $4365\pm120 pc$. For comparison,
\cite{Watkins2015} found a shorter distance, $4150\pm80 pc$, from a dynamical
estimate, while the true distance modulus derived by Woodley et al.~(2012)
from WDs corresponds to a larger distance, $4699\pm176 pc$. We refer the reader
to Table 1 of \cite{Woodley2012} for a list of distance modulus
measurements for 47\,Tuc that have been compiled from the scientific literature. 

By definition, the true distance modulus should be the same regardless of the
filter used for the measurement, but we find a larger value by 0.02-0.03 mag
from the $V$-band than from $F606W$. This could be caused by zero-point errors
in either of the photometric data sets or in the bolometric corrections of
\cite{Casagrande2014}, or both. We note that comparing the cluster sequence in
the {\it HST} ACS $F606W,F814W$ photometry of the central field of 47\,Tuc from
\cite{Sarajedini2007} and our photometry in the outer field
containing V69, as we do in Fig.~\ref{fig:io}, suggests a zero-point difference
of $\sim0.04$ mag (such that $F606W$ magnitudes seem fainter in the central
field). This difference needs to be considered if comparing our distance moduli
from $F606W$ observations to the results of other studies that used the
$F606W$ photometry of \cite{Sarajedini2007} (or \citealt{Milone2012} which employ
the same zero-point). If one were to trust the \cite{Sarajedini2007}
photometric zero-point more, the true distance modulus would be larger by
0.01-0.02 mag when derived from $F606W$ rather than $V$, instead of lower by
0.02-0.03 mag when using our independently measured zeropoint.

Because of the zeropoint differences, we conducted an external check;
specifically, we used the same procedure to calculate the $T_{\rm eff}$
value and distance moduli for the eclipsing binary V40 in the globular cluster
NGC6362 \citep{Kaluzny2015}. That system has
{\it HST} ACS photometry in the $F606W$ and $F814W$ bands from
\cite{Sarajedini2007}, where V40 has ID 2539 in the photometry of NGC6362.
In this case, we found that the true distance modulus is larger by 0.03 mag
in $V$ than in $F606W$, suggesting that this is likely to be the precision
level that can be reached, probably due to the uncertainty of the photometric
zeropoints. Therefore, we adopt as our best estimate of the true distance
modulus $(m-M)_0=13.21\pm0.06$ (random) $\pm0.03$ (systematic).

We emphasize that the procedure we describe and use below is insensitive to the
photometric zeropoints when it comes to deriving the age; i.e., only the
derived distance, but not the age, is affected by these zeropoint issues. 

   \begin{figure*}
   \centering
   \includegraphics[width=14cm]{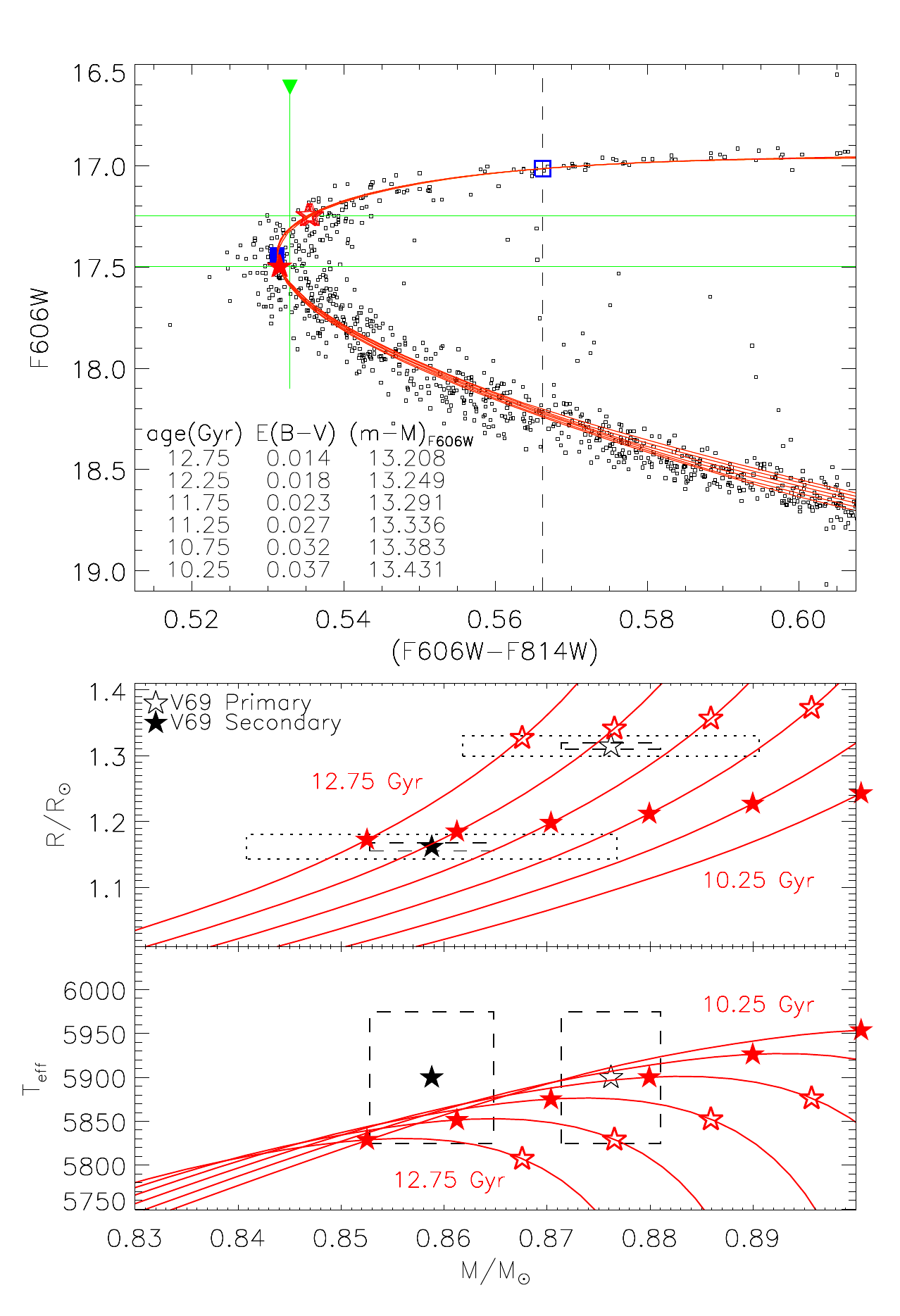}
   \caption{Isochrone fits to the CMD and the components of V69 using the procedure explained in the text. The top panel shows the $F606W,F606W-F814W$ CMD of 47\,Tuc in the outer field where V69 is located. The solid blue square shows location of the cluster turn-off and the open blue square identifies the point on the subgiant branch (SGB) which is 0.035 mag redder than the turn-off. Isochrones with ages from 10.25 to 12.75 are all positioned on the observed CMD so that the point which is 0.035 mag redder than the turn-off of the isochrone matches the corresponding point on the CMD. The green triangle shows the location of the combined light of V69 while the horisontal green lines indicate the magnitudes of the individual components. The positions on the isochrones with these magnitudes are marked with red star symbols and give predictions for the properties of V69 from the given isochrone. 
The lower panels show mass--radius and mass--$T_{\mathrm{eff}}$ diagrams. Black star symbols mark the observed values for the components of V69, while dashed lines and dotted lines indicate $1\sigma$ and $3\sigma$ uncertainties, respectively. Overlaid are the same isochrones as in the top panel. The masses, radii and $T_{\mathrm{eff}}$s at the CMD magnitudes of the V69 components are again marked with red star symbols.}
              \label{fig:procedure}%
    \end{figure*}

\section{Matching model isochrones to the observations}
\label{sec:procedure}
Our procedure, which we demonstrate using Fig. \ref{fig:procedure}, makes use
of the CMD cluster sequence in the colour range from the turn-off colour to a
point that is 0.035 mag redder in $F606W-F814W$. \citetalias{VandenBerg2013} showed
that the shape of isochrones within 0.05 mag of the turn-off colour
is quite insensitive to changes in age, composition, and model physics and
therefore suitable for the age determination of metal-poor clusters
(because any isochrone will provide the same fit to the observations in
the vicinity of the turnoff if the appropriate distance modulus for that
isochrone is adopted).  Since the shapes of isochrones of higher
metallicity, such as those that apply to 47\,Tuc, are somewhat more sensitive
to age, we chose to narrow the colour range that is fitted from
0.05 to 0.035 mag. As can be seen in Fig.~\ref{fig:procedure} we would
obtain even smaller differences between isochrones if we had chosen an even
smaller colour range, e.g., 0.02 mag. However, this could cause contamination
by the binary sequence crossing the SGB; also, our {\it HST} photometry of the outer field
does not have many SGB stars that are approximately 0.02 mag redder than the
turn-off (see Fig.~\ref{fig:procedure}).

We first determined the colour of the turn-off (marked by a solid blue square
in Fig. \ref{fig:procedure}) in the {\it HST} photometry containing V69 by
binning the data in magnitude and then finding the median color in
each bin.  We then derived the $F606W$ magnitude of the point on the SGB that
is 0.035 mag redder than the turn-off (blue open square), by binning the
observed cluster sequence in color and evaluating the mean magnitude in each
bin.  Different choices of bin-size and summing methods affected the $F606W$
magnitude so derived only at the level of $\leq 0.003$ mag.

The green triangle in Fig. \ref{fig:procedure} marks the CMD position of the total light of V69. To obtain the individual apparent magnitudes of the V69 components we adopted the luminosity ratio for the $V$-band as derived by \cite{Thompson2010}. The $T_{\rm eff}$ of the two components is very close to identical \citep{Thompson2010}, and therefore we are only making a very small error in adopting this light ratio also for the $F606W$ filter. Any error in the light ratio and in the assumption of identical $T_{\rm eff}$ of the V69 components will be largely compensated when considering both components together, since a potential overestimate for one component will translate into an underestimate for the other. The green vertical line marks the system colour and the green horizontal lines the $F606W$ magnitudes obtained for the individual components.

In this study, isochrones of a chosen chemical composition but
different ages are first shifted in colour to match the observed
turn-off colour and then adjusted in the vertical direction to match the
magnitude of the point on the SGB that is 0.035 mag redder than the
turn-off.  This procedure, which is equivalent to the isochrone fitting
method employed by VBLC13, gives the distance modulus as a function of
age.  Consequently, the measured distance from, e.g., Sect.~2.6 could
similarly be used to derive an age.  However, our aim here is to invoke
the constraints of radii and $T_{\rm eff}$\ on age separately, and in
combination with the masses, independently of the luminosity and distance
modulus.

The isochrones are interpolated from a grid of stellar models as described by
\cite{VandenBerg2014} and we refer the reader to that paper for details. We
show for each case the nominal reddening, $E(B-V)$, as well as the apparent
distance modulus needed to satisfy the CMD constraints.
We then interpolate the isochrones to the intersections with the measured
$F606W$ magnitudes of the V69 components, marked with red star symbols, to obtain
the corresponding values for mass, radius, and effective temperature, which we
compare to the observed values in the mass-radius and mass--$T_{\rm eff}$
diagrams in the lower panels. Open (solid) star symbols correspond to the
primary (secondary) component.

Since the apparent magnitude of V69 is measured to high precision in exactly
the same photometry that is used for the CMD, the procedure is insensitive to
any zero-point issues with the photometry when it comes to estimating age (but
not distance). This can be understood by considering that a zeropoint offset
would shift everything in the top panel of Fig.~\ref{fig:procedure} vertically
by the same amount and thus leave the isochrone predictions for the masses,
radii, and effective temperatures of V69 unchanged. Furthermore, the V69
components are located in the turn-off region of the CMD, very close to the
SGB in magnitude, so we are only relying on the bolometric corrections to be
trustworthy in a relative sense over very limited magnitude and parameter
ranges to provide the age estimate.

Following this procedure, we adjusted the age of the isochrone in steps 0.5 Gyr
from 10.25 Gyr to 12.75 Gyr, to be able to estimate the best age for a chosen
chemical composition. Fig.~\ref{fig:procedure} shows this for the specific case
for which $[\rm{Fe/H}]=-0.70, [\rm{\alpha/Fe}]=+0.4, [\rm{O/Fe}]=+0.6$, and
$Y=0.25$. The isochrones are ordered in the legend according to their
brightness on the main sequence. In the mass--radius diagram, the measured
values of the V69 components are given with $1 \sigma$ uncertainties shown as
dashed boxes and $3 \sigma$ uncertainties as dotted boxes. The isochrone
comparisons suggest an optimal age for the chosen composition somewhere
between 12.25 Gyr and 12.75 Gyr depending on whether more weight is given to
the masses or to the radii. The isochrone effective temperatures for these ages
are also compatible with our earlier estimate, especially when considering
the fact that the models suggest a lower reddening, in which case our
photometric $T_{\rm eff}$ estimate would also have been lower. It should
be appreciated that the $E(B-V)$ values implied by fits of isochrones to the
turnoff colour are affected by potential systematic errors in the photometry, the model $T_{\rm eff}$\ scale
and/or in the adopted color transformations.  As a result, the numbers listed
for this parameter could well involve appreciable uncertainties.

The example shown in Fig~\ref{fig:procedure} turns out to be an exceptional
case, since for most other reasonable composition choices, we are unable to
match both the masses and the radii close to their observed $1\sigma$ limits
at the same time.

We repeated the above procedure for the $BVI_C$ photometry using the
$V,V-I_C$ and $V,B-V$ fiducial sequences constructed by \cite{Bergbusch2009}
from their photometry. This resulted in self-consistent results for the
apparent distance modulus $(m-M)_V$ between the two colour-planes that is
$\sim0.03$ mag larger than $(m-M)_{F606W}$ just as we found for the apparent
distance of V69 that was derived previously. However, the colour of V69 was
offset from the observed cluster sequence, being bluer than the turn-off in
$V,V-I_C$ and redder than the turn-off in $V,B-V$. This could suggest that our
measured $V$ magnitude is too small by $\sim0.01$ mag, since that would remove
the discrepancy. At the CMD positions of the binary components, the radius is
changing much more rapidly than either mass or $T_{\rm{eff}}$ as a function of
magnitude. Our procedure therefore predicts radii larger by $1 \sigma$ from the $V,V-I_C$
and $V,B-V$ colours compared to $F606W,F606W-F814W$, while the masses and
temperatures are not significantly different. If we were to adopt a 0.01
mag increase to the $V$ magnitude, as suggested by the $BVI_C$ solutions, the
radii, as well as masses and effective temperatures, would agree at the level
of $\lesssim0.1\sigma$. However, the reddening implied by the $V-I_C$ and $B-V$
colour solutions are too high and too low, respectively, relative to the
$F606W-F814W$ solution, suggesting that it is the colours of the fiducial
sequences that are problematic, rather than the colours of V69. Some
evidence for this can be found in the fact that stars in the vicinity
of V69 in a $V,V-I_C$ CMD scatter only on the blue side of the
fiducial sequence, which is constructed from stars occupying a much
larger field. However, the ground-based photometry does not have a large
sample of stars measured to high precision in the close vicinity of V69,
making it difficult to resolve this issue. Because of this, we choose to rely
on the {\it HST} photometry for our model predictions, while keeping in
mind that the ground-based data would predict either identical or larger radii
by $1 \sigma$.   

   \begin{figure*}
   \centering
   \includegraphics[width=19cm]{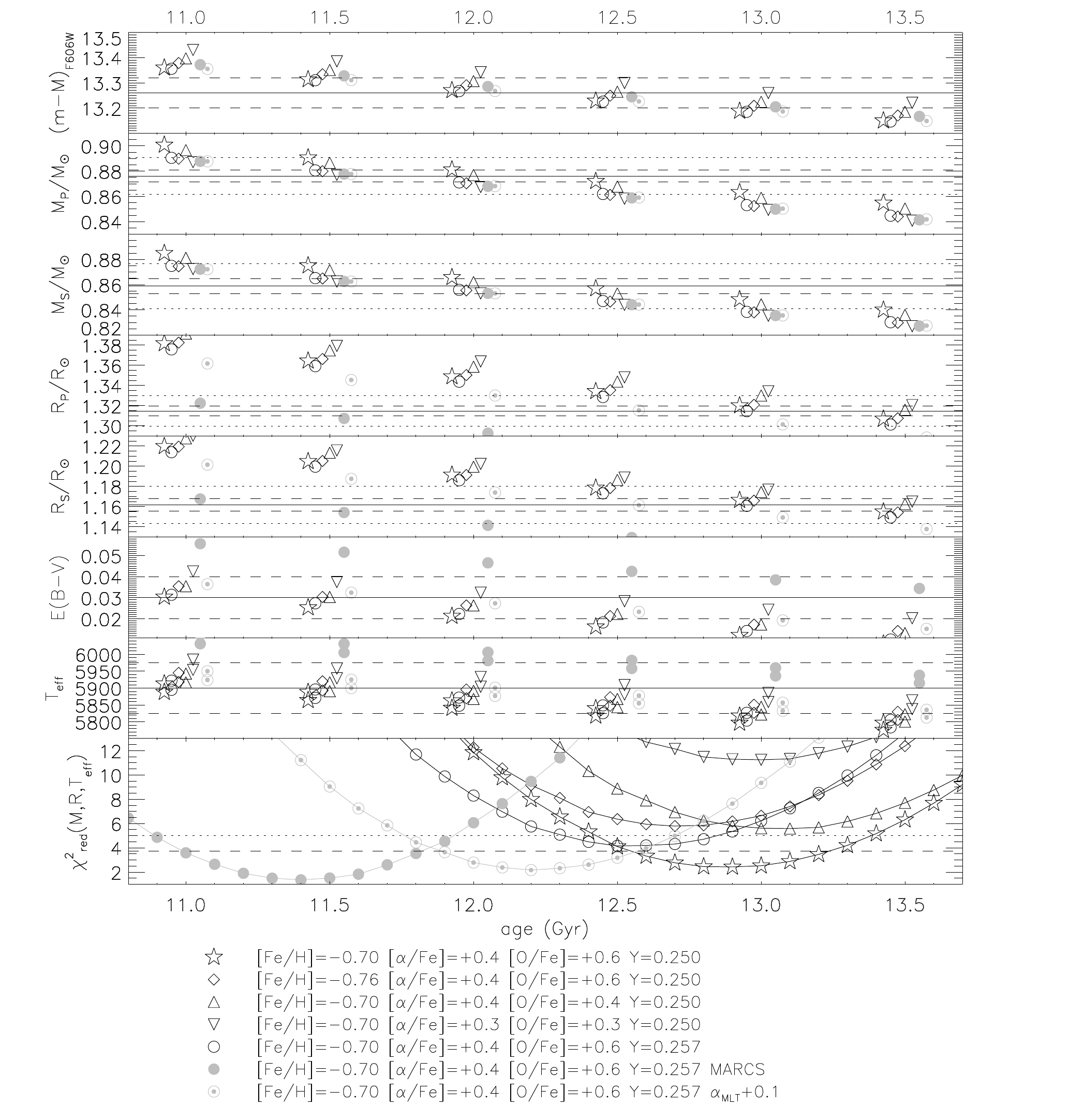}
   \caption{Comparison between measured and predicted values for properties of 47\,Tuc and the components of V69 for different assumptions on composition (black) and model physics (grey). From top to bottom, the panels compare the apparent distance modulus in $F606W$, the masses of the primary and secondary component of V69, the corresponding radii, the nominal reddening, and $T_{\rm{eff}}$ to the model predictions. Solid lines are observed values, dashed (dotted) lines their $1 \sigma$ ($3 \sigma$) uncertainties. Predictions are shown for ages between 11.0 and 13.5 Gyr in steps of 0.5 Gyr. Note that for each age, the prediction from different isochrone assumptions have been slightly shifted along the age axis for visibility purposes. The bottom panel shows the reduced $\chi^2$ value calculated using masses, radii and $T_{\rm eff}$ for the same age range, but with a 0.1 Gyr spacing. In this panel, the dashed and dotted lines are the upper limits for models considered as good or acceptable matches, respectively, according to the criteria given in the text.}
              \label{fig:compare}%
    \end{figure*}
%

   \begin{figure}
   \centering
   \includegraphics[width=9.5cm]{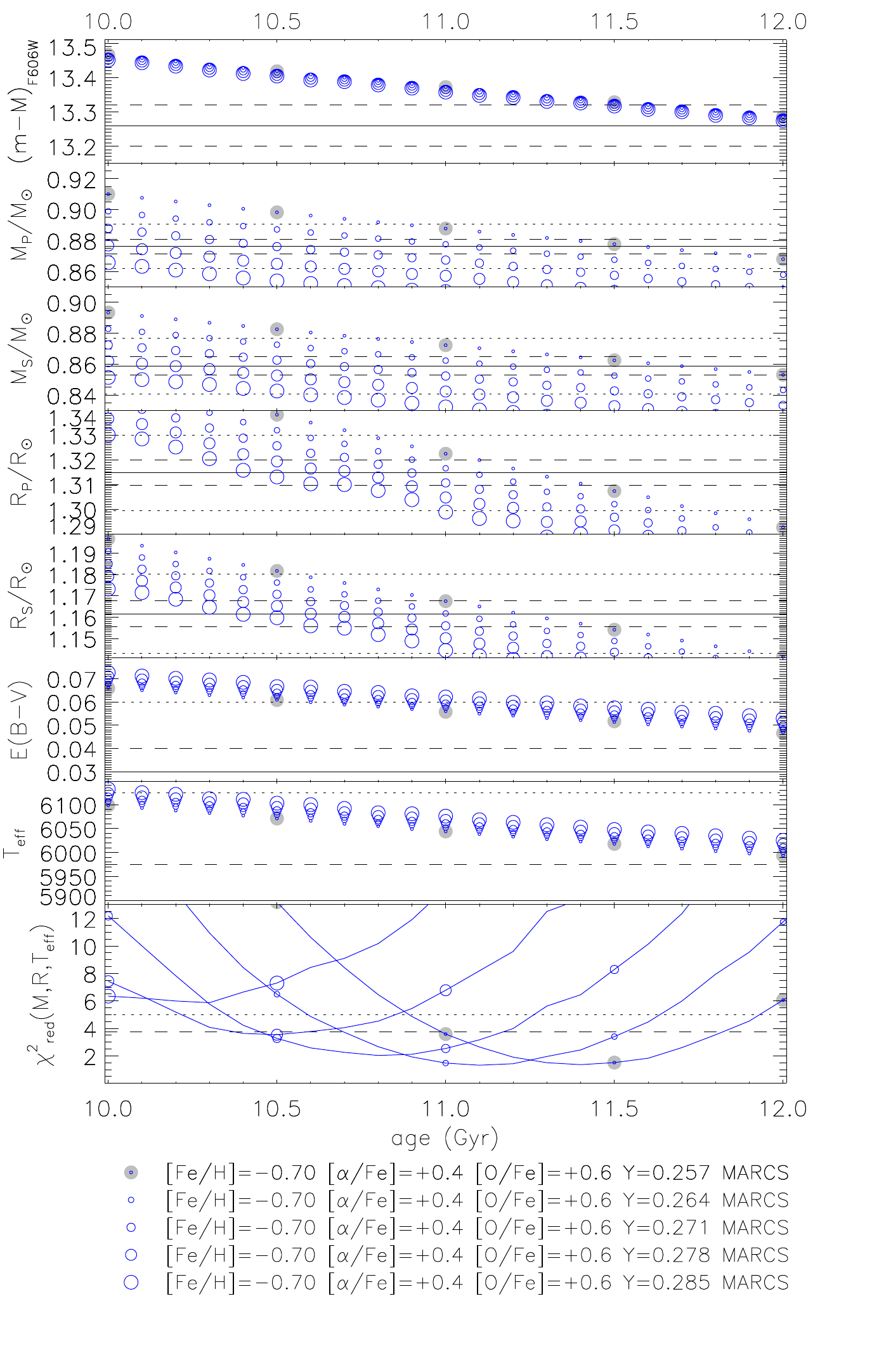}
   \caption{Similar to Fig.~\ref{fig:compare} but showing only isochrone predictions for cases using 1D MARCS atmosphere models as surface boundary conditions, assuming varying amounts of helium mass fraction $Y$. For visibility purposes, only the mean values of the two components are shown in the $T_{\rm eff}$ panel. The grey solid circles are to indicate the corresponding model in Fig.~\ref{fig:compare}.}
              \label{fig:compareY}%
    \end{figure}
%

   \begin{figure}
   \centering
   \includegraphics[width=9.5cm]{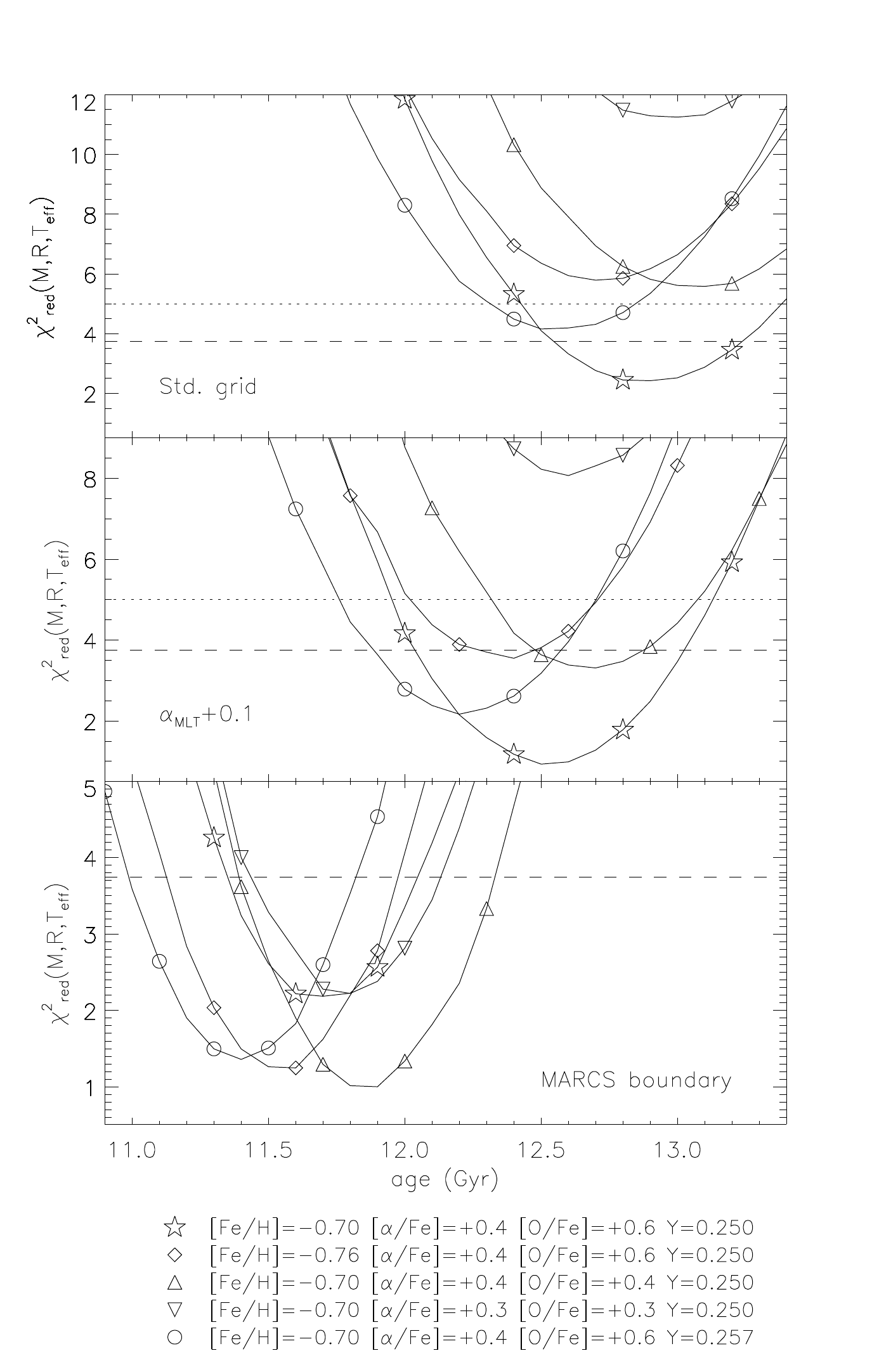}
   \caption{Similar to the bottom panel in Fig.~\ref{fig:compare}, but with the $\chi^2$ in each panel calculated for isochrones with different assumptions for the surface physics. The top panel is for the standard model grid. In the middle panel the isochrones have an increased mixing length parameter. In the bottom panel, 1D MARCS atmosphere models have been used for the surface boundary conditions. Note the different $\chi^2$ ranges in each panel.}
              \label{fig:compare2}%
    \end{figure}
%

   \begin{figure*}
   \centering
   \includegraphics[width=15cm]{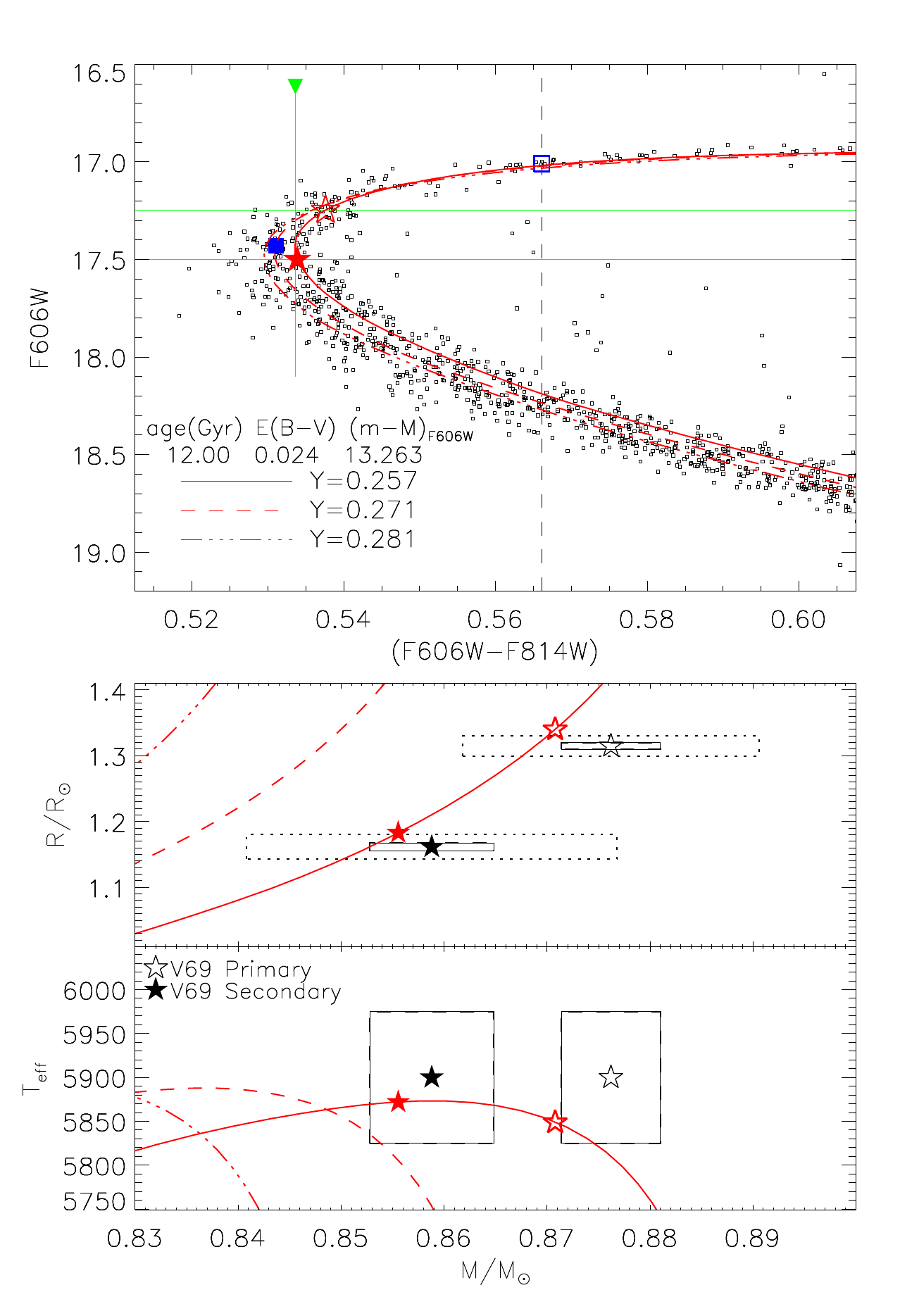}
   \caption{Similar to Fig.~\ref{fig:procedure}, except that in this case only the solid isochrone is taken to represent V69 and a potential helium-poor population, while two others of the same age represent potential populations with higher $Y$ values.}
              \label{fig:dY}%
    \end{figure*}

\section{Results and discussion}

Fig.~\ref{fig:compare} shows the result of repeating the above procedure for various composition choices as given in the figure legend. The case marked with the star symbols corresponds to the example in Fig.~\ref{fig:procedure}. From top to bottom, the figure compares measurements of the apparent distance modulus, the masses of the primary and secondary component of V69, the corresponding radii, the nominal reddening, and $T_{\rm{eff}}$ to the isochrone predictions. Solid lines are observed values, dashed (dotted) lines their $1 \sigma$ ($3 \sigma$) uncertainties. 

We have used the chi-square ($\chi^2$) test to determine which isochrones
provide viable fits to the masses, radii and $T_{\rm{eff}}$s of V69. The bottom panel plots
the reduced $\chi^2$ value from masses, radii and $T_{\rm{eff}}$ of V69 which we explain below. The lines in this panel correspond to a theoretical (dashed) and an empirical (dotted) $3 \sigma$ limit, as determined in the next section.

Conclusions can be drawn from Fig.~\ref{fig:compare} on different levels depending on how well one relies on stellar models to be able to reproduce the correct radii and thus effective temperatures of stars. 
First of all, the distance modulus versus age relations in the top panel can be used, as in the study by \citetalias{VandenBerg2013}, to derive the preferred age
under the assumption of a given distance modulus, e.g. the distance modulus derived in Sect.~\ref{sec:distance}.
Another approach is to consider only the masses. This corresponds to matching the observed mass--luminosity relation of V69. {\it From the mass panels alone it can then be deduced that the age of 47\,Tuc must be $\sim11.75\pm0.5$ Gyr $(1 \sigma)$ and older than 10.75 Gyr at the $3 \sigma$ level if the composition of 47\,Tuc can be represented by any of the isochrones shown}. However, by inter-comparing the various isochrone predictions, it is also evident that younger ages can be reached by adopting a model with a lower [Fe/H] or higher helium mass fraction $Y$. For better constraints on the age the radii must be considered as well.

Examining the radius panels in Fig.~\ref{fig:compare}, it becomes evident
that older ages are obtained by demanding that the stellar models reproduce
both the mass and radius measurements of V69. Almost all of the model radii are
larger than measured by more than three sigma at ages where the masses are in
optimal agreement. The reduced $\chi^2$ values in the bottom panel are
calculated from the masses, radii and $T_{\rm{eff}}$ of both components. The exact expression used is
\begin{equation}
\label{eqn:chi}
\chi^2_{\rm red}=\frac{1}{6}\sum_{_{_{n=1,2}}}{\sum_{_{_{X=M,R,T}}}{\frac{(X_{\rm n,obs}-X_{\rm n,iso})^2}{\sigma_{X_{\rm n,obs}}^2}}}
\end{equation}
with {\it obs} denoting observations and {\it iso} the isochrone predictions while
the summation of $n$ runs over the two components of V69 and the summation of $X$ runs through mass, radius, and $T_{\rm{eff}}$.
The reduced $\chi^2$ minima reveal that
the best isochrone matches are obtained at ages of 12.5 Gyr and older.
The few exceptions to this rule (grey solid circle and double-circle) arise
from changing physical assumptions of the models. We return to those later.   

\subsection{How well do we expect to match V69 properties?}

By definition, the reduced $\chi^2$ values should be close to one if measurements are
independent and errors are random. For measurements of eclipsing
binaries we know that this is not strictly the case due to correlations between the parameters used to derive the masses and radii. 
In addition, the binarity could potentially be causing the evolution of the V69 components to be slightly different from that of single clusters stars, although we expect this effect to be negligible due to the detached configuration (P=29.54 days) and the agreement between the components and the cluster sequence in the {\it HST} photometry. We therefore investigated the mass--radius diagrams for old star clusters where multiple eclipsing systems have been measured in order to estimate what reduced $\chi^2$ value corresponds to an acceptable model. 

For the specific cases of M4 \citep{Kaluzny2013}, NGC6362 \citep{Kaluzny2015}, and NGC6791 \citep{Brogaard2011,Brogaard2012}
we estimated the reduced $\chi^2$ values that one would get by comparing the binary measurements to the isochrones presented in the same studies. 
These reduced $\chi^2$ values turned out to be very close to, or even slightly above, the number which defines the upper 0.1\% probability of the reduced $\chi^2$ distribution for the specific number of parameters. This suggests that either the best models were not found, or uncertainties were not correctly estimated.
Although it cannot presently be ruled out that these higher-than-expected numbers are caused by inaccurate models in the respective studies, we adopt a
conservative approach in the following and define an empirical $3\sigma$ upper limit in addition to the theoretical.
To define the empirical limit we used the two eclipsing systems V18 and V20 in NGC6791 \citep{Brogaard2012} because spectroscopic measurements of $T_{\rm eff}$ are available in addition to masses and radii. We conservatively assumed that an isochrone that matches V20 within $1\sigma$ is correct and then calculated the reduced $\chi^2$ from the masses, radii and $T_{\rm eff}$s of V18. This yielded a reduced $\chi^2$ value of 5.0 (calculated using eqn.~\ref{eqn:chi} with 1, 3, and 1 $\sigma$ for the mass, radius and $T_{\rm{eff}}$ of the primary and correspondingly 3, 3, and 1 $\sigma$ for the secondary as determined from the mass-radius diagrams). We will use this as a conservative empirical $3\sigma$ upper limit in the following. 
For comparison, the theoretical $3\sigma$ upper limit, which we chose to be the $\chi^2$ value above which there is a 0.1\% probability of reaching by chance, is $\chi_{\rm red}^2$=3.743 for the 6 parameters\footnote{$\chi_{\rm red}^2$=3.743 is obtained from $\chi^2_{\rm red}=\frac{1}{6}\times\chi^2$ for 6 parameters and $\chi^2=22.46$ calculated by the Python SciPy command scipy.stats.chi2.isf(0.001, 6)} (2 masses, 2 radii, 2 $T_{\rm eff}$). We leave it to the reader to decide which of the limits to use. In the following we denote isochrones as good (acceptable) representatives of the observations if they have reduced $\chi^2$ values below 3.743 (5.0) and mark both values in figures showing reduced $\chi^2$ estimates. Any isochrone resulting in a $\chi^2$ value above 5.0 is not considered to be an acceptable representation of the observations.

\subsection{Composition effects in the standard model grid}

As seen in Fig.~\ref{fig:compare} the predictions from the isochrone marked by
star symbols are the only ones from the standard model grid (black symbols)
that result in a good match according to the $\chi^2$ criterion. The other
cases shown are chosen such that only one of the composition variables ([Fe/H],
$[\rm{\alpha/Fe}]$, [O/Fe], or $Y$) are changed relative to the first (as is common practice, a change to [Fe/H] refers to a change of all elements by the same relative amount as Fe). They
show that acceptable matches can be obtained at slightly higher values of $Y$,
or slightly lower values of [Fe/H] or [O/Fe] relative to the reference case. 

The effects on parameters of changing composition are additive to a very good
approximation which reveals that e.g. increasing $Y$ {\it and} reducing [O/Fe]
relative to the ``best-fit" model (star symbols) will not result in an
acceptable match. The case with $[\rm{\alpha/Fe}]=+0.3$ does not result in an
acceptable fit. This is caused almost exclusively by [O/Fe] following
$[\rm{\alpha/Fe}]$ in this case, which shows that low-oxygen models are
incapable of representing V69. 

For the standard model grid, only isochrones that assume a high oxygen
abundance, $[\rm O/Fe]=+0.6$, are able to fit the masses and radii at an
acceptable level, unless a lower value of $Y$ is adopted.
We emphasize that even though we varied the [O/Fe] value of the isochrones it is in fact the sum of log$\epsilon
({\rm C+N+O})$ which matters \citep{Rood1985,VandenBerg2012}. For an $\alpha$-enhanced mixture oxygen is however much more abundant than carbon and nitrogen.

Due to the Na--O anti-correlation observed in globular clusters, the model [O/Fe] should not be compared to
the mean abundance derived for an ensemble of stars, but rather to the upper
limit of values found (of course taking into account measurement uncertainties).

As far as we are aware, the only spectroscopic evidence for [O/Fe] $\sim+0.6$ in 47\,Tuc
is the study by \cite{Koch2008}. Other studies have found [O/Fe]$\sim+0.50$ at the high-oxygen end of the Na--O anti-correlation (e.g. \citealt{Dobrovolskas2014,Thygesen2014}) but there are also reports of a lower value of [O/Fe]$\sim+0.3$ \citep{Carretta2009,Cordero2014}.

Since our models assume the abundances given by Asplund et
al.~(2009) for the reference solar mixture, the same absolute oxygen abundance
would result from a lower $[\rm O/Fe]$ if one adopted the
\cite{Grevesse1998} solar abundances for the isochrone grid instead. However,
\cite{Marino2016} measured log$\epsilon({\rm C+N+O})=8.48\pm0.07$ from stars on the
SGB of 47\,Tuc with the \cite{Asplund2009} solar abundances as the reference
abundances, just as we have assumed. We calculated the corresponding
log$\epsilon({\rm C+N+O})$ values of our isochrones using the description and
tabulated values in \cite{VandenBerg2014}. At [Fe/H] $=-0.70$ log$\epsilon
({\rm C+N+O})$ is 8.495 for $[\rm O/Fe]=+0.4$ and 8.659 for $[\rm O/Fe]=+0.6$.
Thus, the measurements of \cite{Marino2016} suggest that we should use
$[\rm O/Fe]=+0.4$. This puts a some tension between our isochrone comparisons using the standard model grid
and the measurements by Marino et al. and others who find values of [O/Fe]$\leq+0.4$.

It is interesting in the same context that the adoption of a lower iron
abundance, $[\rm Fe/H]=-0.76$, and $[\rm O/Fe]=+0.6$ resulted in acceptable
solutions only if it is assumed that $Y = 0.250$ (or less), which is very
close to current best estimates of the primordial helium abundance --- even 
though $[\rm Fe/H]=-0.76$ is within the uncertainties of many recent
spectroscopic studies \citep{Koch2008,Thygesen2014,Lapenna2014,Cordero2014,Carretta2009}.
However, while this could be an indication that the true metallicity of 47\,Tuc
is closer to $[\rm{Fe/H}]=-0.68$ as found by \cite{Johnson2015} and previously
by \cite{Carretta2004} and others, or even be an indication that globular
clusters do not follow a helium enrichment law $\Delta Y/\Delta Z$, this
depends too much on other details of the stellar models to be a significant
result.
This is demonstrated in the next section.

While our tests of chemical abundance effects indicate a preference for
relatively high [Fe/H] and [O/Fe] values and a low helium abundance, they also
show that the age is not very sensitive to the exact composition. All of the
acceptable cases from the standard model grid have $\chi^2$ minima at ages between 12.5 and 13 Gyr and no
age younger than 12.3 Gyr is accepted by our reduced $\chi^2$ criterion of 5.

\subsection{Mixing length and surface boundary conditions}
\label{sec:alternative}
In contrast to our finding of an age of 12.3 Gyr or higher, \cite{Hansen2013}
derived an age of $9.9\pm0.7$ Gyr, at $95\%$ confidence, using measurements of
white dwarfs in 47\,Tuc. They supported this determination by 
reanalyzing V69, the eclipsing binary also used in the present study, for
which they obtained an age of $10.39\pm0.54$ Gyr (without describing the details of how it was obtained, unfortunately). In order to check whether
variations in uncertain model assumptions could result in a younger age we
therefore constructed additional isochrones. 

To be specific, model grids were generated in which (i) the solar-calibrated
value of the mixing-length parameter $\alpha_{\rm{MLT}}$ was increased by 0.1, or (ii) 1D MARCS
model atmospheres were employed as surface boundary conditions.  (In the
standard grids, the pressure at T=$T_{\rm eff}$ was calculated by integrating the
hydrostatic equation on the assumption of the empirical Holweger-M\"uller (HM)
T-tau relation.)

Predictions from these isochrones are shown as grey filled circles and
double-circles in Fig.~\ref{fig:compare}, which may be compared with the
corresponding model from the standard grid, shown as a black open circle. As
can be seen, the luminosity and mass are almost unaffected in these alternative
models, while the effective temperatures are higher at the expense of smaller
radii. This is completely in line with expectations, since changes to the
mixing length and surface boundary conditions affect the size of the model star,
and the effective temperature is forced in the opposite direction in order to
satisfy $\frac{L}{L_{\odot}}=(\frac{R}{R_{\odot}})^2\times(\frac{T_{\rm eff}}{T_{\rm eff,\odot}})^4$,
where the luminosity $L$ is determined deep within the star and therefore
is only marginally affected.

As evidenced by the reduced $\chi^2$ minima the isochrones with increased
mixing length or alternative surface boundary conditions are able to match the
V69 observations better than most, and as well as any, cases from the standard
model grid. Importantly, they both do so at significantly younger ages although
they are still $\gtrsim 11$ Gyr. A natural question would then be whether
enhancing one of the effects or adding both effects to the same isochrone could
produce a good match at an even younger age. However, the predicted masses of V69, which are
unaffected by the surface assumptions, are already above their $3 \sigma$ upper
limits at 10.75 Gyr and much worse at younger ages. To reproduce the 
observed masses at younger ages, adjustments to the composition would
be needed (e.g. increased $Y$ or decreased [\rm{Fe/H}]), but that would
increase further the predicted temperatures, which are already
approaching their upper limits using the alternative boundary conditions. 

We show in Fig.~\ref{fig:compareY} the same kind of comparison between isochrones and observations as in Fig.~\ref{fig:compare} but for isochrones assuming MARCS atmosphere conditions, and with varying amounts of helium mass fraction $Y$. Since in this case the lower age limit is mainly determined by $T_{\rm eff}$ due to the change to the model temperature scale, we argue that the theoretical $\chi_{\rm red}^2=3.743$ limit should be used, given that our $T_{\rm eff}$ measurements already include a potential systematic error (Sect.~\ref{sec:teff}). As can be seen, an age as young as 10.4 Gyr is allowed by the theoretical $3\sigma$ limit, although only close to the $3 \sigma$ upper limits of both the radii and $T_{\rm eff}$ of V69 while requiring a high oxygen and helium content {\it and} a change to the model $T_{\rm eff}$
scale that is as large as the most extreme case we show (adopting MARCS model
atmospheres as surface boundary conditions without redoing the solar
calibration). We argue below against this possibility. 

Since changes to the mixing length and surface boundary conditions have a
significant effect on ages that are derived using our approach, we encourage
new and on-going efforts to incorporate the results from 3D hydrodynamical
atmosphere simulations into stellar models \citep{Magic2015,Mosumgaard2016,Aarslev2017}. At present, the predicted
metallicity (and $\log g$ and $T_{\rm eff}$) dependence of the mixing length
from 3D hydrodynamical simulations \citep{Magic2015} suggest that the mixing
length of the V69 components and turn-off stars in 47\,Tuc should be decreased
by 0.04--0.08 relative to the solar value, instead of increased by 0.1 as in
our example. While that would make implied ages older by 0.25--0.50 Gyr in
Fig.~\ref{fig:compare}, it would also make isochrone fits to the 
observations of V69 worse if using the standard model grid. We therefore expect that the corresponding changes
to the surface boundary conditions, which are not yet available in a form that
can be easily implemented in 1D stellar models, will have a compensating and
likely larger effect in the other direction. If we reverse the effects of our
$\alpha_{\rm{MLT}}+0.1$ case to become instead a $\alpha_{\rm{MLT}}-0.1$, as
approximately predicted by the 3D hydrodynamical atmosphere simulations
(remember that effects are additive to a good approximation), and also add the
effect of using the 1D MARCS model atmospheres as surface boundary conditions
as a poor-man's prediction of the effects of a 3D model atmosphere, we 
expect to end up with predictions close to, but slightly hotter than those for the
$\alpha_{\rm{MLT}}+0.1$ case that will match very well all of our observables of
V69 including its $T_{\rm{eff}}$ at an age close to 11.8 Gyr.

In order to do so, the models with MARCS atmosphere boundary conditions should
satisfy the solar constraints. This can be accomplished in several ways, the
simplest being a readjustment of $\alpha_{\rm{MLT}}$. However, since the MARCS
models do not provide a good representation of the solar atmosphere
\citep{VandenBerg2008, Pereira2013}, alternative procedures might be preferred,
such as a rescaling of the pressure in the model atmospheres. This procedure
was investigated by \cite{VandenBerg2008} and again by \cite{VandenBerg2014}
for models that take diffusion of helium into account. For the latter case,
which is representative of our models, it was found that low metallicity models 
with MARCS surface boundary conditions are hotter than those using the HM
boundary conditions, even when the pressure in the MARCS atmospheres has been
rescaled so that a $1 M_{\odot}$, [Fe/H]$=0$ model is almost identical for the
two cases (see their Fig. 4). Thus, 1D MARCS atmosphere boundary conditions
suggest a somewhat hotter $T_{\rm eff}$ scale than the standard model grid.

In fact, it is even possible that MARCS model atmospheres without any ad hoc
correction to the photospheric pressure provide a more realistic $T_{\rm eff}$
scale at low metal abundances than those in which the boundary pressure has been
modified so as to satisfy the solar constraint.  Under such assumptions, which corresponds to the MARCS cases of the present paper, the
evolutionary tracks at low $Z$ are even hotter (see Fig.~4 by 
\citealt{VandenBerg2014}). Although a solar abundance, $1 M_\odot$ track is also
shifted to higher temperatures, one cannot conclude that models with HM
atmospheres ``represent those of metal-deficient stars better than standard
MARCS models just because the latter are problematic for metal-rich dwarfs"
\citep{VandenBerg2008}.  That is, it is entirely possible that stellar models
which employ HM boundary conditions become progressively too cool as the
metallicity decreases.  Fortunately, on-going advances in theory can be 
expected to help resolve this issue.  For instance, \cite{Pereira2013} have shown
that the temperature structure predicted by 3D model atmospheres reproduces
the observed center-to-limb variation of the continuum intensity of the Sun
extremely well.

In any case, since we use the same isochrones as \cite{VandenBerg2014} (based on
HM boundary conditions), we can make use of their comparisons to observations of
local subdwarfs (see their Figs.~12 and 13) to make an assessment of the
accuracy of the $T_{\rm eff}$ scale predicted by these models. If we consider
only the subdwarfs with [Fe/H]$\geq-1.5$, they are on average 30~K hotter than
the model predictions, whereas our $T_{\rm eff}$ for V69 is hotter by
$\sim50$--75\ K. Their subdwarfs with [Fe/H]$\leq-1.5$ are on average cooler
than the model predictions but that could be telling us that stellar models
should assume that $\alpha_{\rm{MLT}}$ decreases towards lower metallicity, as
predicted by 3D atmosphere simulations \citep{Magic2015}. Thus, the nearby
subdwarfs provide reasonably good support for the predicted $T_{\rm eff}$
scale of our standard model grid, though higher temperatures by about 50\ K cannot be ruled out.
Because a 100 K increase would cause all of the subdwarfs considered by
\cite{VandenBerg2014} to be cooler than model predictions, we consider
such a high temperature offset to be an upper limit.

Our inaccurate knowledge in modeling surface conditions reduces the significance
of our earlier inferences about the composition of 47\,Tuc.
Fig.~\ref{fig:compare2} shows the reduced $\chi^2$ values as a function of age for different assumptions on the surface conditions, corresponding to different $T_{\rm eff}$ scales. As seen, shifting from the standard model grid to a slightly hotter $T_{\rm eff}$ scale represented by the isochrones with $\alpha_{\rm{MLT}}+0.1$ results in improved matches for all composition choices without affecting the order of preference. However, at the even hotter $T_{\rm eff}$ scale represented by the isochrones with MARCS atmosphere boundary conditions, there is even a change to the order of composition preference. 
Thus, if this $T_{\rm eff}$ scale turns out to be correct then earlier conclusions relating to composition are no longer valid.

Regardless of the possible changes to the composition that are allowed by adopting alternative
boundary conditions, we still find no acceptable isochrone match which results
in an age younger than 10.4 Gyr. Furthermore, this lower limit to the age is only reached close to the upper $3\sigma$ limits of $T_{\rm eff}$ and radii
while requiring an increase to both the observed and the model $T_{\rm eff}$ scale of $\gtrapprox100\,K$ and a high value of $Y\sim0.278$.

Given this and the fact that current stellar models from different researchers
produce nearly identical results when the same physics is adopted (see, e.g.,
Fig.~1 in \citealt{VandenBerg2016}), we conclude that, unless there are some
significant shortcomings in current stellar models that we have yet to realize,
the age of 47\,Tuc must be strictly larger than 10.4 Gyr and most likely significantly larger. In our investigation, $11.8^{+1.6}_{-1.4}$ Gyr covers the total range of acceptable solutions within a very conservative $3\sigma$ limit; 11.8 Gyr is also the age that is obtained if the model $T_{\rm eff}$ scale yields a temperature for V69 that is identical to our measured value.

\subsection{Effects of multiple populations}

As is well-known by now, globular clusters habour multiple populations, and 47\,Tucanae is no exception, with the first photometric evidence presented by \cite{Anderson2009}. Although the exact meaning of multiple populations and how they arise is still under investigation it has been inferred that there are star-to-star variations in the helium content \citep{diCriscienzo2010,Milone2012,Salaris2016}. Since the brightness of the SGB is insensitive to the helium content our procedure is largely unaffected by this and is in principle able to put some constraint on the helium content of the stars - at least those that belong to the same population as V69. 

The star-to-star helium mass fraction variation in 47\,Tuc has been inferred from different methods to be somewhere between $\Delta Y=0.015$ \citep{Milone2012} and $\Delta Y=0.03$ \citep{diCriscienzo2010,Gratton2013}. To demonstrate how such a variation in helium content will affect the cluster sequence we show in Fig.~\ref{fig:dY} our isochrone for $Y=0.257$, representing the lowest-$Y$ population. If we assume that V69 belongs to the lowest-$Y$ population and that there is an intrinsic helium difference between populations then an isochrone for the helium content of the lowest-$Y$ population should be used to match the mass--radius diagram of V69 and set the distance in accordance with our procedure. However, the observed cluster sequence in the CMD should not be matched by this same isochrone, but by some average of isochrones with a distribution of helium contents that represent the cluster. We use isochrones of the same age and distance, but higher values of $Y$, 0.271 and 0.281, as representatives of the higher-$Y$ population in Fig.~\ref{fig:dY}. 

The effective temperature at the turn-off is slightly different for different helium contents. If V69 represents a population with a lower helium content than the mean, then the turn-off colour of that population is redder than the mean turn-off colour. Therefore, the point on the isochrone which is 0.035 mag redder than the turn-off (of the isochrone) should be matched to the point on the observed SGB which is redder than the observed turn-off by 0.035 mag plus the colour-difference from the mean turn-off colour due to helium difference. From Fig.~\ref{fig:dY}, 0.0025 mag seems representative of the extra shift. If applied, the inferred radii are $\sim 0.6 (0.5) \sigma$ smaller for the primary (secondary) component, thus alleviating slightly some of the tension of the previous matches to the isochrones of the standard model grid where the predicted radii are usually too large. The fact that the isochrone points representing the V69 components both become slightly redder than the combined light in Fig.~\ref{fig:dY} is not significant, since 0.0025 mag is within the photometric measurement uncertainty. The implied changes to the radii would suggest that V69 belongs to the lowest-$Y$ population when using the standard model grid, where the radii would give slightly better isochrone matches than the previous ones due to the smaller predicted radii. Assuming instead a high-$Y$-population scenario for V69 would would cause a colour-shift in the other direction and imply larger radii, worsening those isochrone matches. Also, within the standard model grid there are no acceptable solutions for high helium contents, suggesting again a lowest-$Y$ population membership for V69.
The indications that the eclipsing binary belongs to the lowest-$Y$ population are in accordance with expectations that most binary stars likely belong to the lowest-$Y$ population (see \citealt{Bedin2013} and references therein). However, the potential offset to the $T_{\rm eff}$ scale discussed earlier could affect these inferences significantly. 

The lower $3\sigma$ age limit of 10.4 Gyr derived in Sect.~\ref{sec:alternative} was only reached at high $Y=0.278$. This suggests a high-$Y$ population scenario for V69 which would, as mentioned above, cause the predicted radii to increase by $\sim 0.6 (0.5) \sigma$ thus pushing the lower age limit older by another 0.2 Gyr while lowering the maximum allowed $Y$ value. A low-$Y$ scenario would shift the lower age limit even older.

As seen in Fig.~\ref{fig:dY}, the multiple-population scenario will broaden
the main sequence, and change the median fiducial sequence, which
becomes some average of the representative isochrones depending on the
distribution of helium contents among the stars.  As an example, we assume in
Fig.~\ref{fig:dY} that the total variation in helium is $\Delta Y=0.024$
(from 0.0257 to 0.281) with a distribution that is skewed slightly towards the
higher-$Y$ population as inferred by \cite{Milone2012}. The resultant mean
helium content would then be close to $Y=0.271$ and thus the isochrone for
that helium content is the one that should match the mean fiducial sequence
in Fig.~\ref{fig:dY}, with a width that is defined by the other two isochrones
(plus observational scatter).  This isochrone matches the complete part of
the isochrone within 0.035 mag of the turnoff in colour, and on the main sequence even outside the box we usually consider. 
Since all the single-population scenarios have the isochrones running along the cool edge of the main sequence, the CMD seems to prefer a multi-population scenario with V69 as part of the lowest-$Y$ population although we caution the reader not to put too much reliance on such a subtle effect. 

It would be useful to measure the magnitudes of V69 in some
of the population-sensitive {\it HST} filters, and/or to do detailed abundance
analysis on disentangled spectra to determine which of the populations it
belongs to. The latter would also allow a direct measure of the effective
temperatures of V69, which if done to high precision would enhance the
constraint on the $T_{\rm eff}$ scale, allowing a more precise upper limit on
$Y$ and a stronger constraint on the age.

\section{Comparison to WD cooling sequence ages}

As mentioned earlier there is some systematic uncertainty in either of the
photometries and/or the bolometric corrections, causing the true distance
modulus to be larger by 0.02 mag when derived using the $V$ filter compared
to $F606W$. Here, we adopt the result from the $V$ band for the following
discussion, noting that a shorter distance modulus, as inferred from $F606W$
would yield slightly older WD ages.

We derived earlier an apparent distance modulus in $V$ of $13.30\pm0.06$ as the
best estimate from our procedure. The corresponding true distance modulus is
$13.21\pm0.06$ if the reddening is $E(B-V)=0.03$. \cite{Woodley2012} and
\cite{Hansen2013} derived true distance moduli from the white dwarfs in 47\,Tuc,
obtaining $13.36\pm0.08$ and $13.32\pm0.09$, respectively, for an adopted
$E(B-V)=0.04$. While these are compatible with ours to within the uncertainties,
the latter estimates are higher than ours by $2-2.5\sigma$. This suggests that
the young age derived by \cite{Hansen2013} is caused at least partly by the
relatively large distance modulus that they adopted --- especially since
\cite{GarciaBerro2014} obtained an age close to 12 Gyr from the white dwarfs
in 47\,Tuc on the assumption of a true distance modulus of $13.20$
(Garcia-Berro, private comm.). This latter study is thus in better
agreement with ours, both in terms of age and distance. Similarly, the study
of the white dwarf cooling sequence by \cite{Campos2016} is in agreement
with our findings. Making a rough extrapolation from the two examples for
47\,Tuc in their section 4.1.2 to our apparent distance modulus,
$(m-M)_V=13.30$, would yield an age of $11.43_{-0.2}^{+0.4}$ Gyr, where
we adopt similar but slightly larger uncertainties than they did due to our
rough extrapolation.

Examining the details of the distance moduli derived by \cite{Woodley2012}
and \cite{Hansen2013}, it turns out that their results depend on the assumed
mass of the white dwarfs at the top of the WD cooling sequence. If our preferred true distance modulus, 13.21,
at a reddening of $E(B-V) = 0.03$, is used in equation (8) in \cite{Woodley2012},
we obtain $M_{\rm WD}=0.583 M_\odot$ for the mass of the white dwarfs in
47\,Tuc, $\sim10\%$ larger than $M_{\rm WD}=0.53 M_\odot$ that was
assumed by \cite{Woodley2012} and $0.525 M_\odot$ by \cite{Campos2016} for the mass of the white dwarfs entering the cooling sequence in
47 Tuc (their Fig. 11).


\section{Potential consequences for relative GC ages}

As mentioned in the introduction, 47\,Tuc was used as an anchor-point for the
metal-rich clusters in the study of GC ages by \citetalias{VandenBerg2013}. They
adopted for this system the distance modulus derived from the eclipsing cluster
member V69, as measured by \cite{Thompson2010}, in order to establish the
stellar mass at the lower envelope of the observed HB.  This reference mass
turned out to be $\approx 0.70 M_\odot$, implying a maximum mass loss of
$\approx 0.20 M_\odot$ during the preceding evolution --- in reasonably good
agreement with current determinations of the mass loss along the giant branch of
47 Tuc (e.g., see \citealt{Origlia2007}, \citealt{Salaris2016}).  By assuming
that the same mass loss occurs in all other metal-rich GCs, their reference ZAHB
masses, the corresponding absolute magnitudes, and the implied distance moduli
could be determined.  Ages then followed from the application of the
$\Delta V^{\rm HB}_{\rm TO}$ method.\footnote{This procedure involves relatively
small uncertainties because, at the red end of a metal-rich horizontal branch,
a variation in mass has no more than a small effect on the predicted luminosity.
For instance, a $0.75 M_\odot$ ZAHB model for $Y=0.257$ and [Fe/H] $= -0.76$ is
only 0.020 mag brighter than one for $0.70 M_\odot$ and the same chemical
abundances.  Thus, the assumption of a somewhat higher or lower upper limit to
the mass loss would have only minor consequences for the distance moduli and
ages derived by \citetalias{VandenBerg2013}.} Since our present work allows for a
rather broad range of possible ages and corresponding distance moduli for
47\,Tuc, we elaborate here on the potential consequences of our results for the
relative ages of metal-rich GCs.

The stellar models employed by \citetalias{VandenBerg2013} adopted the solar metal
abundances given by \cite{Grevesse1998} as the reference mixture, with
enhancements to the $\alpha$-elements that were derived by \cite{Cayrel2004} at
low metallicities. Because different $\alpha$-elements are enhanced by
different amounts (e.g., [O/Fe] $= +0.5$, whereas [Mg/Fe] $= +0.3$; for
a detailed listing, see Table 1 by \citealt{VandenBerg2012}), this resulted
in the net value of $[\alpha$/Fe] $= +0.46$.  Since recent spectroscopic studies
generally report [$m$/H] and [$m$/Fe] values, for individual metal $m$, with respect to the solar
abundances tabulated by \cite{Asplund2009}, we used in the present study the
new grids of evolutionary tracks and isochrones based on this reference
mixture computed by \cite{VandenBerg2014} for $-0.4 \le$ [$\alpha$/Fe] $\le
+0.4$, this time assuming that the abundances of all of the $\alpha$-elements
vary together.  Additional grids that allow for variations of the oxygen
abundance relative to that implied by the selected value of [$\alpha$/Fe] have
also been generated. (They will be described in detail and made generally
available in a forthcoming publication.)

The combination of differences due to the adopted reference solar abundances
and the specific abundance patterns between the study of \citetalias{VandenBerg2013}
and our preferred solutions turns out to be small and would result in older
ages by only 0.25 Gyr. However, when adding the photometric zeropoint offset of 0.04--0.06 mag in $F606W$ between our field with
V69 and the inner field as measured by \cite{Sarajedini2007} (cf. Sect.~\ref{sec:calibration} and Fig.~\ref{fig:io}) and used by
\citetalias{VandenBerg2013}, the distance moduli inferred by our standard model grid
would be shorter by 0.04--0.06 mag compared to the value of $(m-M)_{F606W}=13.34$
adopted by \citetalias{VandenBerg2013} (see the black stars and open circles in the upper
panel of Fig.~\ref{fig:compare} close to the corresponding $\chi^2$ minima in
the bottom panel). On the other hand, our preferred metallicity is also higher
than the [Fe/H] value adopted by \citetalias{VandenBerg2013}, while the helium
mass-fraction is the same or smaller. These differences would cause
the HB models to be fainter, which would partly compensate for the shorter
distance modulus.  

As a result, the inferred mass at the lower envelope of the 47\,Tuc HB would
remain close to, or just slightly less massive than, the reference mass derived
by \citetalias{VandenBerg2013}, if redetermined from our standard model grid, while
the age would be older by 0.25--0.75 Gyr.  Depending on whether or not a similar
photometric zeropoint problem is present in the data for the other metal-rich
clusters and on the metallicities that are adopted for them, there could be a
similar revision to their ages.  In this case, the overall appearance of the
age--metallicity relation determined by \citetalias{VandenBerg2013} would remain
unaltered for the metal-rich group of clusters, while there would be a small
overall shift of 0.25--0.75 Gyr towards older ages relative to those derived
for lower metallicity clusters.  While it is possible that the age of 47 Tuc is
as high as 12.5--13 Gyr, making it approximately coeval with the most
metal-deficient GCs (see, e.g., the studies of M\,3, M\,15, and M\,92 by
\citealt{VandenBerg2016}), such high ages seem highly improbable for most of
the other metal-rich clusters.  According to \citetalias[see their
Fig.~14]{VandenBerg2013}, most of them appear to have formed 10.75--11.25 Gyr
ago.  To increase these estimates by 1.25--1.75 Gyr would require either reduced CNO abundances by $\sim 0.5$ dex,
or the adoption of smaller distance moduli by 0.12--0.18 mag, which would make the
observed HBs considerably fainter than ZAHB models for the appropriate
metallicities. 

Note that, if one accepts the indications in Sect.~\ref{sec:alternative} that a
hotter model temperature scale such that V69 has $T_{\rm eff}$ very close to
5900 K is more appropriate, then we would obtain nearly perfect agreement with
the \citealt{VandenBerg2013} results for the 47\,Tuc age and distance modulus (on
the zero-point of \citealt{Sarajedini2007}) but not for its metallicity.  A
higher [Fe/H] value (i.e., $\gtrapprox -0.70$, for which there has been some
observational support over the years; e.g., \citealt{Carretta2004},
\citealt{Alves-Brito2005}, \citealt{Wylie2006}) is suggested by the binary
constraints.  It is obviously extremely important to discover and analyze
eclipsing binaries in other metal-rich GCs in order to determine whether they
also suggest a revision to the cluster [Fe/H] scale, which could have some
ramifications for the slope of the age--metallicity relation.

\section{Summary, conclusions and outlook}

We have presented for the first time a self-consistent isochrone fitting method to
globular cluster CMDs and an eclipsing binary member. 47\,Tuc was found to
be older than 10.4 Gyr at the $3\sigma$ level, and likely significantly older.
For our best estimate of the model $T_{\rm{eff}}$ scale, we obtain an age of
$11.8$ Gyr (assuming [Fe/H] = $-0.7$, [alpha/Fe] = $+0.4$, [O/Fe] = $+0.6$), which is favored by the
observed mass-luminosity relation of the binary.  On the other hand,
if equal weight is given to the observed radii as to the observed
masses, and the $T_{\rm{eff}}$ scale of the standard model grid is trustworthy,
then the best age estimate rises to 12.5 Gyr, or even 12.75-13.0 Gyr if $Y$ is very close to the primordial He abundance.

Consistency between our best estimate age and that from the white dwarf cooling sequence is automatically obtained since adopting the distance modulus from our analysis in a white dwarf analysis results in a WD age ($11.43_{-0.2}^{+0.4}$ Gyr using results from \citealt{Campos2016} or $\sim12$ Gyr from \citealt{GarciaBerro2014}) in agreement with the age from our analysis.

Our work indicates a preference for a multiple-population scenario with the
eclipsing binary V69 belonging to the lowest-$Y$ population with high oxygen,
[O/Fe]$\sim+0.60$, and low helium content, $Y\sim0.250$. We inferred an upper
limit to the helium mass fraction of $Y\lesssim0.278$ for V69. 

Our conclusions regarding the composition and the precise age of 47\,Tuc are very dependent on the helium content and the model $T_{\rm{eff}}$ scale, which is 50-75 K cooler than our best estimate but within measurement uncertainties. Therefore, the potential impact on the GC relative ages and age-metallicity relation derived by \citetalias{VandenBerg2013} remains uncertain until our procedure can be repeated for a number of GCs and/or geometrical GC distances can be determined.

The analysis of 47\,Tuc can be improved in several ways. As already mentioned
by \cite{Thompson2010}, the uncertainty of the mass measurements of the V69
components can be decreased by 33\% by doubling the existing number of radial
velocity measurements, thereby leading to stronger constraints. Furthermore, if
enough spectra of sufficient S/N can be gathered, then the effective
temperatures of the V69 components can be directly derived from disentangled
spectra, improving also the constraint on their effective temperatures
which is important for determining the correct distance and $T_{\rm{eff}}$
scale for the models. Such an analysis could also reveal which population V69
belongs to by measuring, e.g., its Na and O abundances.

Perhaps even more interesting is the possibility of adding an additional
eclipsing member to the analysis. A very suitable eclipsing SB2 system, E32,
quite similar to V69 has already been identified and suggested for further
observation and analysis by \cite{Kaluzny2013}. Having two additional points in
the mass--radius diagram along, with two more distance measures, would tighten
the constraints significantly.

Although one of the strengths of the method is that the age estimate is
insensitive to the true distance, there is still much additional benefit in
knowing the true distance to the cluster, especially given the significant
span of distances derived from various methods, or even from our method alone
when varying the chemical composition and/or input physics. Therefore, a
potential precision distance measurement from the Gaia mission
\citep{Gaia2016,Pancino2013, Pancino2017}, would be of benefit to tighten the constraints  ---
not only on the distance but also on the effective temperature scale of the
models when combined with the binary measurements.

The full potential of the method presented will be reached when applied to a
larger sample of globular clusters with eclipsing binary members, since this
will allow much more precise relative GC ages than presently available and
enable a detailed study of the $T_{\rm{eff}}$ scale of current and future
stellar models. Due to the large effort that is needed to find and analyse
eclipsing systems in globular clusters, this has only been done for a small
number of these objects so far, all within the CASE project \citep{Kaluzny2015}.
These known systems will benefit from a reanalysis using our method, but for
some cases, {\it HST} photometry for both the cluster and the eclipsing
system --- a key ingredient for the method --- is not available. Thus, future
efforts should focus on obtaining {\it HST} photometry for these known systems
while finding and analysing eclipsing binaries in more globular clusters.

\section*{Acknowledgements}

We wish to thank Luca Casagrande for useful discussions on interstellar reddening. We thank P. Stetson, J. Kaluzny and E. Garcia-Berro for quick and useful responses to inquiries related to the present work. 

Funding for the Stellar Astrophysics Centre is provided by The Danish National Research Foundation (Grant DNRF106). The research was supported by the ASTERISK project (ASTERoseismic Investigations with SONG and Kepler) funded by the European Research Council (Grant agreement no.: 267864).
KB acknowledges support from the Villum Foundation and the Carlsberg Foundation.  A Discovery Grant from
the Natural Sciences and Engineering Council of Canada supported the contribution 
to this project by DAV. APM acknowledges support by the Australian Research Council through the Discovery Early Career Researcher Awards DE150101816.




\bibliographystyle{mnras}
\bibliography{mnras_brogaard_47Tuc} 








\bsp	
\label{lastpage}
\end{document}